\documentclass[12pt]{article}
  
\usepackage{amsmath,amssymb}

\usepackage{subfigure}

\usepackage{xspace}  
\usepackage{epsfig}  
  
\usepackage{graphicx}               

\newcommand{\nc}{\newcommand}  

\nc{\beq}{\begin{equation}}  
\nc{\eeq}{\end{equation}}  
\nc{\beqa}{\begin{eqnarray}}  
\nc{\eeqa}{\end{eqnarray}}  
\nc{\bea}{\begin{eqnarray}}  
\nc{\eea}{\end{eqnarray}}  
\nc{\ra}{\rightarrow}  
\nc{\lsim}{\begin{array}{c}\,\sim\vspace{-21pt}\\< \end{array}}  
\nc{\gsim}{\begin{array}{c}\sim\vspace{-21pt}\\> \end{array}}

\nc{\LL}{L}  
\nc{\vv}{\tilde{v}}  
\nc{\GG}{\widetilde{G}}  
\nc{\MM}{\ensuremath{\mathcal{M}}}  
\nc{\UU}{\ensuremath{\mathcal{U}}}  
\nc{\ZZ}{\ensuremath{\mathcal{{\cal{Z}}}}}  
\nc{\Mu}{\ensuremath{M_{u}}}
\nc{\Md}{\ensuremath{M_{d}}}
\nc{\Mtu}{\ensuremath{\tilde{M}_{u}}}
\nc{\Mtd}{\ensuremath{\tilde{M}_{d}}}

\title{  
\Large  
\textbf{TeV Scale Singlet Dark Matter}\vspace*{1.0cm}   
\author{\large  
\textbf{Eduardo Pont\'{o}n$^{a}$}
and  
\textbf{Lisa Randall$^{b}$},  
\\
$^a$\normalsize\emph{Department of Physics, Columbia University,}\\
\normalsize\emph{538 W. 120th St, New York, NY 10027, USA} \\  
$^b$\normalsize\emph{Jefferson Laboratory of Physics, Harvard University,}\\
\normalsize\emph{Cambridge, Massachusetts 02138, USA}
}
\date{}}  
\begin{document}  
\setcounter{page}{0}  
\maketitle  
\vspace*{1cm}  
\begin{abstract}  
It is well known that stable weak scale particles are viable dark
matter candidates since the annihilation cross section is naturally
about the right magnitude to leave the correct thermal residual
abundance.  Many dark matter searches have focused on relatively light
dark matter consistent with weak couplings to the Standard Model.
However, in a strongly coupled theory, or even if the coupling is just
a few times bigger than the Standard Model couplings, dark matter can
have TeV-scale mass with the correct thermal relic abundance.  Here we
consider neutral TeV-mass scalar dark matter, its necessary
interactions, and potential signals.  We consider signals both with
and without higher-dimension operators generated by strong coupling at
the TeV scale, as might happen for example in an RS scenario.  We find
some potential for detection in high energy photons that depends on
the dark matter distribution.  Detection in positrons at lower
energies, such as those PAMELA probes, would be difficult though a
higher energy positron signal could in principle be detectable over
background.  However, a light dark matter particle with
higher-dimensional interactions consistent with a TeV cutoff can in
principle match PAMELA data.
\end{abstract}  
\newpage  
  
\setcounter{page}{1}

\baselineskip18pt    

\section{Introduction}
\label{sec:intro}  

Dark matter has received a lot of attention of late as new dark matter
searches ramp up.  Of particular interest is the increasing capacity
to detect dark matter in both direct and indirect channels.  The
latter rely solely on dark matter annihilation, which is nice in that
it doesn't assume any particular type of interaction with the Standard
Model and furthermore the annihilation rate is generally connected to
the annihilation cross section responsible for the current dark matter
abundance. 

Given the importance of dark matter searches and our lack of knowledge
as to the true nature of dark matter, it makes sense to explore the
range of possibilities and what their implications would be for
current and future detectors.  In this paper we will consider singlet
dark matter candidates with mass of order one to a few ${\rm TeV}$.
We assume a $Z_2$ symmetry that prevents any operator allowing decay
and therefore ensuring stability.  This is perhaps the simplest dark
matter candidate there can be.  In fact, such a possibility has been
previously considered in
Ref.~\cite{Silveira:1985rk,McDonald:1993ex,Burgess:2000yq,Davoudiasl:2004be},
but in a lower mass region.  In this paper we concentrate on the
remaining allowed mass range, of order one to a few TeV, which
phenomenologically is also a viable possibility.  We concentrate on
some novel scenarios that arise in a framework with a low cutoff
scale.

Although we mostly take an agnostic approach about the source of this
dark matter, we also focus on TeV scale particles that arise in a
theory with a TeV cut-off scale.  Such a scenario can occur for
example in the RS framework~\cite{Randall:1999ee}.  See
also~\cite{MarchRussell:2008yu} for a study of heavy DM in a
supersymmetric theory with a relatively low cutoff.

In this paper we show the range of allowed parameters giving the right
relic density and then consider whether such dark matter has any
chance of being detected.  We find that annihilation into photons
might provide a visible signal at high energy gamma ray detectors such
as HESS or VERITAS, particularly if higher-dimension operators are
present.  We also consider more model-dependent scenarios in which
annihilation into positrons can also occur.  We show the signal can
exceed background with reasonable assumptions, but most likely not in
the PAMELA range for a TeV-scale mass.

On the other hand, we observe that a dark matter candidate of about
100 GeV whose dominant decay mode involves direct positron emission,
such as can occur with a higher-dimension operator suppressed by the
TeV scale, matches PAMELA data quite nicely.

\section{Singlet Dark Matter}
\label{sec:relic}  

We start by discussing the relic density computation for a thermally
produced Standard Model singlet.  We consider first a renormalizable
four-dimensional theory.  This analysis would of course also apply to
a nonrenormalizable theory so long as the renormalizable coupling of
the singlet scalar to a Higgs dominates annihilation, including a
five-dimensional theory with a brane-bound scalar or any
five-dimensional theory where the higher-dimension operators are
suppressed.

We then consider a more exotic possibility that could in principle
give rise to a detectable positron signal.  We will see this scenario
is unlikely to explain the PAMELA data, although it could
give rise to a detectable signal in the high-energy positron range.

\subsection{Thermal Relic Abundance for a Singlet}
\label{sec:BraneFields}  

We assume a singlet field $\Phi$ protected by a discrete $Z_2$
symmetry $\Phi \rightarrow -\Phi$ in a nonrenormalizable theory with a
TeV cutoff scale, $\Lambda$.  Without any additional fields, the only
\textit{renormalizable} operator that involves SM fields is
\beqa
{\cal L} \supset  \frac{1}{2} \lambda \, \Phi^{2} H^{\dagger} H~,
\label{4DPhiH}
\eeqa 
where $H$ is the Higgs doublet and $\lambda$ is a dimensionless
coupling.  Such an operator can  arise in an RS scenario
for either IR brane-localized or bulk scalars $\Phi$.  For an IR
brane-localized scalar (assuming the Higgs is also IR localized), the
corresponding operator is
\beqa
{\cal L}_{5} \supset -\delta(L-y) \frac{1}{2} \lambda \, \Phi^{2} H^{\dagger} H~.
\label{PhiHInteraction}
\eeqa
For a bulk scalar $\Phi$ the operator Eq.~(\ref{4DPhiH})
can be induced from a non-renormalizable operator (to be discussed in
the following subsection).  If the cutoff is at the TeV scale the
effective coupling $\lambda$ can easily be of order one, so that the
following analysis applies.

The interaction in Eq.~(\ref{4DPhiH}) can lead to the direct
self-annihilation of $\Phi$ particles into a pair of Higgses, and
also, if the annihilations occur after the electroweak phase
transition, into pairs of SM gauge bosons and fermions through
$s$-channel Higgs exchange.  When the $\Phi$ mass is much larger than
the Higgs mass, the direct annihilation into Higgses dominates (this
includes annihilation into the Goldstone modes, hence the $W_{L}W_{L}$
and $Z_{L}Z_{L}$ channels).  Annihilation into two Higgses in the
limit that $M_{\Phi} \gg v_{\rm EW}$ (with $v_{\rm EW}$ the Higgs VEV)
gives in the non-relativistic regime
\beqa
\langle \sigma_{\Phi\Phi \rightarrow HH} v \rangle &\approx& 
\frac{\lambda^{2}}{16\pi m^{2}_{\Phi}}~,
\label{braneAnnXS}
\eeqa
where $v$ is the relative velocity of the annihilating particles, and
the brackets denote thermal averaging.

Notice that other annihilation channels have to proceed through
operators suppressed either by the cutoff scale $\Lambda$ or by a loop
factor.  As we will argue in the next subsection, when the $\Phi$
particles propagate in the bulk of an RS scenario those channels might
be relevant (and could even dominate depending on couplings) in the
total self-annihilation cross section, and therefore in the
determination of the relic density.  However, for a conventional
four-dimensional scalar (or for a brane-localized $\Phi$ in an RS
scenario\footnote{The operators discussed in
Subsection~\ref{sec:BulkFields} vanish for a brane localized $\Phi$
since always one of the chiralities of any bulk fermion $\Psi$
vanishes on the brane.}) all other channels are expected to give a
relatively small contribution when the $\Phi$ mass is less than the
cutoff scale.  For instance, annihilation into SM fermions would
proceed through operators that also involve the Higgs field, of the
form $\Phi^{2} H \bar{\psi}_{1} \psi_{2}$, and are suppressed at least
by order $(v_{EW}/\tilde{\Lambda})^{2}$, where $v_{EW}$ is the Higgs
vacuum expectation value and $\tilde{\Lambda} = \Lambda \, e^{-kL}$ is
the warped down cutoff scale if in a 5D warped framework, or more
generally the cutoff scale of the 4D theory.  Decays into SM gauge
bosons are also expected to be subdominant so long as the DM mass is
less than the cutoff, even if the cutoff scale is low, and will be
discussed in Subsection~\ref{sec:photons} in the context of DM
indirect signals [see Eqs.~(\ref{XS-photons}) and (\ref{XS-Zphotons})
and ensuing discussion].

\begin{figure}[t]
\centerline{ 
\includegraphics[width=0.45 \textwidth]{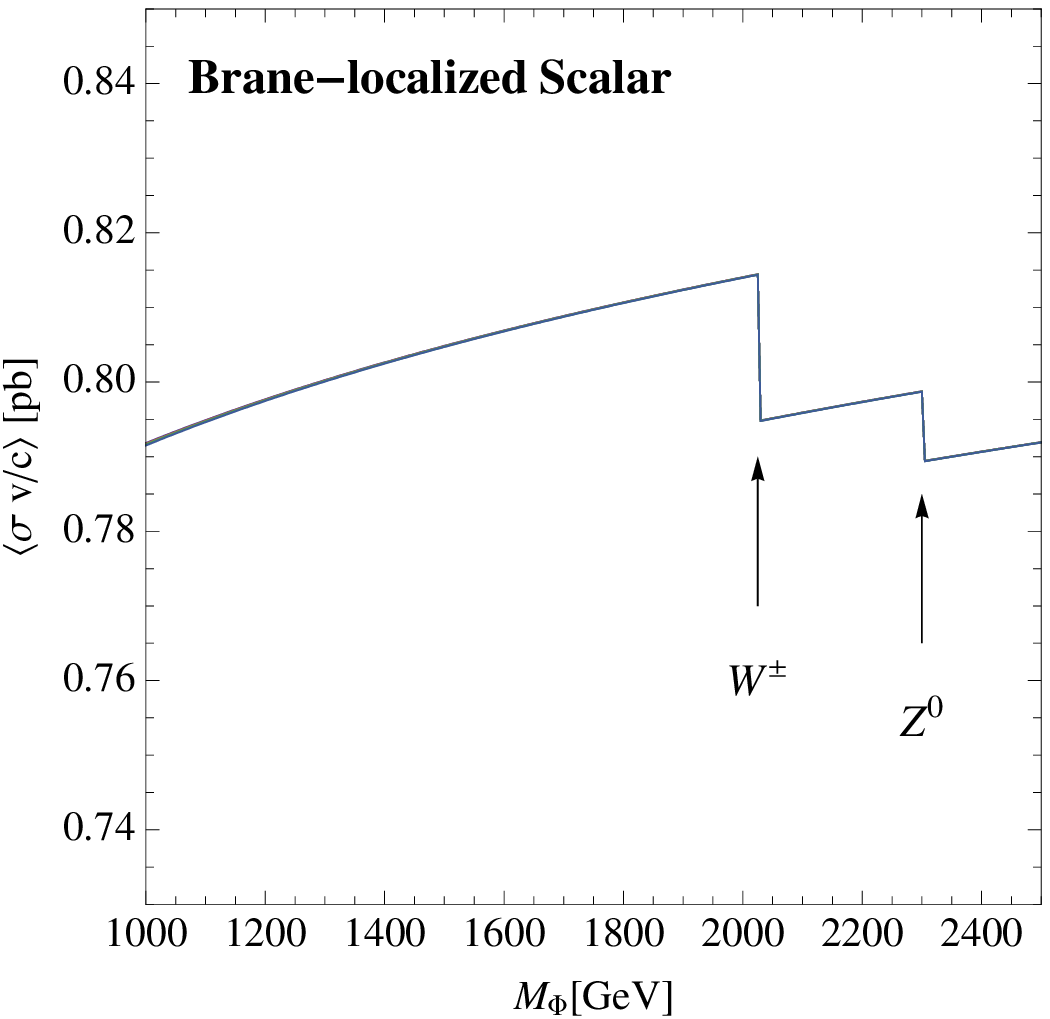}
\hspace*{1cm}
\includegraphics[width=0.432 \textwidth]{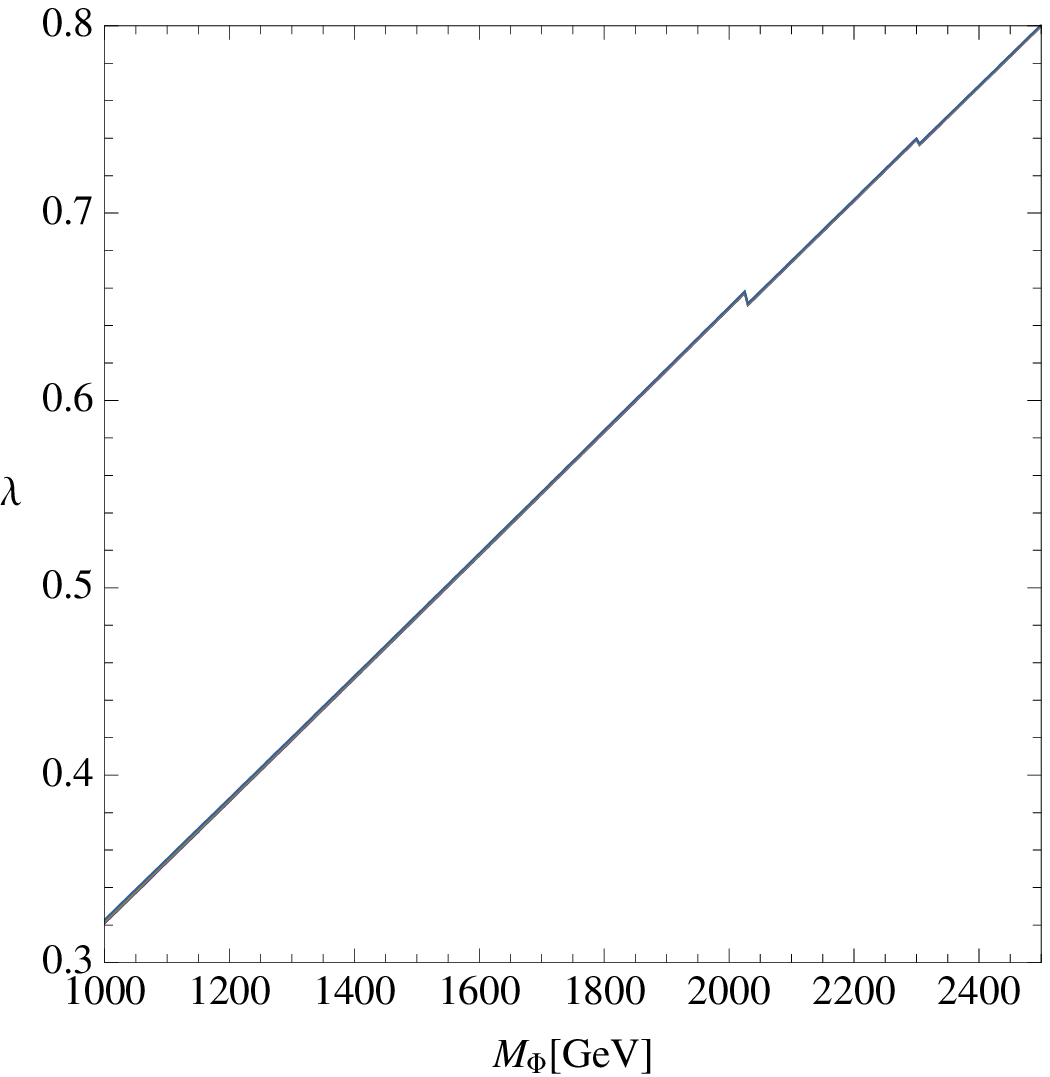}
} 
\caption{Left panel: annihilation cross section, $\langle
\sigma_{\Phi\Phi \rightarrow HH} v \rangle$, for a brane-localized
scalar in the non-relativistic regime as a function of $M_{\Phi}$,
imposing the WMAP constraint on the DM relic density.  The arrows
indicate the points where the freeze-out temperature ($\sim
M_{\Phi}/25$) crosses the $W^{\pm}$ and $Z^0$ thresholds.  Right
panel: the corresponding coupling $\lambda$, defined in
Eq.~(\ref{4DPhiH}), as a function of $M_{\Phi}$.}
\label{fig:branescalar}
\end{figure}

Under the assumption that the DM candidate is heavy (say $1~{\rm TeV}$
or so) and is thermally produced, the DM relic abundance is controlled
by Eq.~(\ref{braneAnnXS}).  Taking into account only the annihilation
into Higgses through the operator Eq.~(\ref{4DPhiH}), and
requiring that the observed DM abundance is completely accounted for
by $\Phi$ particles, we can determine the coupling $\lambda$ as a
function of $M_{\rm \Phi}$ from the WMAP constraint $\Omega_{DM} h^2
\approx 0.11$~\cite{Komatsu:2008hk} and
\beqa
\Omega_{DM} h^2 &\approx& \frac{1.04 \times 10^9~{\rm GeV^{-1}}}{M_P}
\frac{x_F}{\sqrt{g_\ast}} \frac{1}{\langle \sigma v \rangle}~,
\label{Omegah2}
\eeqa
where $M_P \approx 1.22 \times 10^{19}$ GeV is the Planck mass,
$x_F=M_{\Phi}/T_F$, with $T_F$ the freeze-out temperature, $g_\ast$ is
the effective number of relativistic degrees of freedom at freeze-out,
and $\langle \sigma v \rangle$ is the thermally averaged annihilation
cross section times relative velocity in units of ${\rm GeV^{-2}}$.
We have assumed that the $\Phi$ particles are thermally produced and
remain in thermal equilibrium until freeze-out, which requires a
coupling $\lambda > 10^{-8}$~\cite{Burgess:2000yq}.  For masses
$M_{\Phi}$ in the few TeV range, $\lambda$ is always of order unity,
as shown in the right panel of Fig.~\ref{fig:branescalar}, so the
above assumption is self-consistently satisfied.

The required (non-relativistic) cross section as determined from the
WMAP constraint to be $\langle \sigma v/c \rangle \approx 0.8~{\rm pb}$
and is shown in the left panel of Fig.~\ref{fig:branescalar} as a
function of $M_{\Phi}$, where the (weak) dependence of $x_{F}$ on the
cross section and the effective number of relativistic degrees of
freedom $g_\ast$ is included.\footnote{The conversion factor from
${\rm GeV^{-2}}$ to ${\rm pb}$ is $0.3894\times 10^{9}~{\rm
GeV^{2}\,pb}$.  Also, to convert the above cross section from $\rm pb$
into units of ${\rm cm^3\,s^{-1}}$ [the CGS units for $\langle \sigma
v \rangle$] one should multiply the number in $\rm pb$ by
$(10^{-36}~{\rm cm^{2}}) \, c \approx 3\times 10^{-26}~{\rm
cm^{3}\,s^{-1}}$.} Throughout the range of interest we have $x_{F}
\approx 25$, while $g_\ast$ is of order $90$.  The above simple
picture is rather generic for a stable TeV scale scalar field
whenever the effects of higher-dimension operators can be neglected.
However, when the scale suppressing the higher-dimension operators is
near the TeV scale, other more exotic scenarios are possible.  Such a
situation, though less likely, could arise within the RS framework,
and will be illustrated with a bulk scalar in the next subsection.
  
\subsection{Bulk Singlet Dark Matter}
\label{sec:BulkFields}  

Another natural possibility in a 5D warped background is that the DM
candidate arises as the lightest KK mode of a bulk scalar.  In order
to be concrete, and simply for illustration purposes, we will assume
in the following that there is a bulk SM singlet scalar obeying
$(-,+)$ boundary conditions (Dirichlet on the UV brane, Neumann on the
IR brane).  In this case, the mass of the lightest KK mode is
determined by only two dimensionless parameters, and can be easily
below those of the gauge KK resonances (say around 1 TeV), as
discussed in more detail in Appendix~\ref{sec:scalars}.  We also
assume that the SM fermions and gauge fields arise from bulk fields.

The couplings to (an IR localized) Higgs field proceed now through the
higher-dimension operator
\beqa
-\delta(L-y) \frac{\lambda^{\prime}}{2\Lambda} \, \Phi^{2} H^{\dagger} H~,
\label{BulkPhiHInteraction}
\eeqa
where $\Lambda$ is the cutoff scale and $\lambda^{\prime}$ is a
dimensionless coupling.  After KK reduction, this induces a coupling
of $\Phi$s to Higgses similar to the one discussed in the previous
subsection with the identification $\lambda = \lambda^{\prime}
f^{2}_{\Phi^{(1)}}/(\Lambda L) \approx \lambda^{\prime} (2k/\Lambda)$.
Here $f_{\Phi^{(1)}} \approx \sqrt{2kL}$ is the $\Phi^{(1)}$
wavefunction evaluated on the IR brane, where $\Phi^{(1)}$ is the
lightest scalar KK mode (the DM candidate).  If this channel dominates
the self-annihilation cross section, the relic density computation
proceeds in exactly the same way as in the case of a brane-localized
scalar discussed in the previous subsection.  As was mentioned there, the
observed relic abundance requires an effective 4D coupling $\lambda$
of order one.  Notice that for a bulk scalar, in spite of the
suppression $k/\Lambda$, this is easily consistent with the NDA bound
$\lambda^{\prime}\lsim 24\pi^{3}$ \cite{Chacko:1999hg}; in fact, for
$k/\Lambda \sim 1/10$ the fundamental coupling $\lambda^{\prime}$ is
well into the perturbative regime, so that the computation is under
theoretical control.

It is possible, however, that channels other than the annihilation
into Higgses are important or even dominate, which could in principle
differentiate between brane and bulk dark matter candidates.  This
scenario requires the value of $\lambda^{\prime}$ well below its NDA
value with other couplings closer to what NDA would suggest.  Since we
are taking an agnostic attitude and are interested primarily in
potential signatures and ways to identify the various possible
scenarios, we consider this possibility next.

For this analysis it is
useful to rewrite the annihilation cross section into Higgses as
\beqa
\sigma_{\Phi\Phi \rightarrow HH} v &\approx& \frac{\lambda^{\prime2}}{4\pi \tilde{\Lambda}^{2}} 
\left( \frac{\tilde{k}}{M_{\Phi}} \right)^{2}~,
\label{BulkAnnXSHiggs}
\eeqa
where $\tilde{\Lambda} = \Lambda\,e^{-kL}$ is the warped down cutoff
(of order a TeV) and similarly for $\tilde{k} = k\,e^{-kL}$.
Now consider operators involving a 5D fermion field (giving rise to a
SM fermion as its zero-mode), for instance
\beqa
\frac{\lambda_{\psi}}{2\Lambda^{2}} \Phi^{2} \, \overline{\Psi} \Psi~,
\label{BulkPhiPsiInteraction}
\eeqa
where $\Psi$ is the bulk fermion and $\lambda_{\psi}$ is a
dimensionless coupling.  This operator leads to the annihilation of
$\Phi$ particles into a SM fermion and one of its KK resonances, e.g.
\beqa
\frac{\lambda_{\psi} \eta}{2\tilde{\Lambda}(\Lambda L)} 
(\Phi^{(1)})^{2} \left[ \overline{\psi}^{(1)} \psi^{(0)} + 
\overline{\psi}^{(0)} \psi^{(1)} \right]~,
\label{BulkPhiKKPsiInteraction}
\eeqa
where $\psi^{(1)}$ is the first KK mode of the bulk fermion $\Psi$,
and $\psi^{(0)}$ is its zero mode (with a well-defined chirality).
The effective 4D coupling depends on the various extra-dimensional
profiles through
\beqa
\eta = \frac{1}{L} \int^{L}_{0} \! dy \, e^{k(y-L)} 
f^{2}_{\Phi^{(1)}} f_{\psi^{(1)}} f_{\psi^{(0)}}~,
\label{wavefunctionEnhancement}
\eeqa
where all the wavefunctions are normalized as in
Eq.~(\ref{KKnormalization}) of Appendix~\ref{sec:scalars}.

Assuming that the channel $\Phi\Phi \rightarrow \psi^{(1)}
\overline{\psi}^{0}$ is open, i.e. $M_{\psi^{(1)}} + m_{\psi^{(0)}}
\leq 2M_{\Phi}$, the corresponding annihilation cross section
is~\footnote{To simplify notation we will refer to the DM candidate
$\Phi^{(1)}$ simply as $\Phi$.}
\beqa
\sigma_{\psi^{(1)} \bar{\psi}^{(0)}} v &=& \frac{N_{c}\lambda^{2}_{\psi} 
\eta^{2}}{16\pi\tilde{\Lambda}^{2}(\Lambda L)^{2}} \, 
\frac{(s - M^{2}_{\psi^{(1)}})^{2}}{M_{\Phi} s^{3/2}}~,
\label{XSFermions}
\eeqa
where $N_{c} = 3$ for quarks while $N_{c} = 1$ for leptons and, for
simplicity, we neglected the zero-mode mass $m_{\psi^{(0)}}$.  In the
non-relativistic limit one has $\sigma_{\psi^{(1)} \bar{\psi}^{(0)}} v
= a + b \, v^2 + \cdots$, with
\beqa
a =
\frac{N_{c}\lambda^{2}_{\psi} \eta^{2}}{8\pi \tilde{\Lambda}^{2} (\Lambda L)^{2}} \, 
\left(1 - y \right)^{2}~,
\hspace{1cm}
b =
\frac{N_{c}\lambda^{2}_{\psi} \eta^{2}}{64\pi \tilde{\Lambda}^{2} (\Lambda L)^{2}} \, 
\left(1 + 2 y - 3 y^{2} \right)~,
\label{abFermions}
\eeqa
where $y = M^{2}_{\psi^{(1)}}/4M^{2}_{\Phi}$.  The related process
$\sigma_{\psi^{(0)} \bar{\psi}^{(1)}} v$ is given by the same
expression.

The magnitude of the annihilation cross sections into fermions depends
strongly on the localization of the fermion zero-mode through the
parameter $\eta$ of Eq.~(\ref{wavefunctionEnhancement}).  Recall that
the fermion zero-mode wavefunctions are proportional to $e^{(1/2 -
c_{f})k y}$, where $c_{f}$ parametrizes the 5D fermion mass in units
of the AdS curvature scale $k$.  The massive KK mode wavefunctions are
all strongly localized near the IR brane.  There are a number of
distinct scenarios according to how flavor is generated:
\begin{enumerate}
\item The SM fermion mass hierarchies arise from the exponential
wavefunction localization and the overlap with an IR localized Higgs
field.  This scenario has the advantage that both calculable flavor
changing effects (from KK gluon exchange), as well as non-calculable
effects from flavor changing non-renormalizable operators, are
significantly
suppressed~\cite{Huber:2000ie,Agashe:2004ay,Fitzpatrick:2007sa,Perez:2008ee}.
One expects the third generation quarks (most likely the right-handed
top) to couple most strongly to $\Phi$.

\item All fermions share the same parameter $c_{f}$, and are localized
close to the IR brane ($c_{f} < 1/2$), so that their couplings to
$\Phi$ are sizable.  Somewhat more generally, EW precision constraints
allow different localization parameters for different fermions so long
as those fermions having identical quantum numbers have nearly the
same $c_{f}$ (when IR localized; otherwise we are in scenario 1
above).  In these scenarios, as-yet unspecified flavor-violating
interactions would be necessary to explain the fermion mass
hierarchies, while not generating dangerous FCNC effects from
higher-dimension operators suppressed by the TeV scale.

\item Fermion mass hierarchies arise from localization in the extra
dimension but the Higgs field is located on or near the UV brane (for
example, if the Higgs mass is stabilized by supersymmetry (SUSY) and SUSY
breaking is connected to the IR scale).  In this case, the lightest
fermions would be localized closer to the IR brane and have the
largest couplings to $\Phi$. 
\end{enumerate}
Among the fermion channels, $\Phi$ annihilates dominantly into the
fermions closest to the IR brane, since the $\Phi$ wavefunction is
localized near the IR brane.  To calculate the annihilation rate, we
need to estimate the expected size of these couplings, which can then
be compared to the couplings to Higgses discussed above, or to the
annihilation into gauge bosons (see Subsection~\ref{sec:photons}).

For a fermion localized near the IR brane (localization parameter
$c_{f} < 1/2$, but not very close to $1/2$), one finds $\eta \sim
(1/5kL) (\sqrt{2kL})^{3} \sqrt{(1-2c_{f})kL} \approx \sqrt{\frac{1}{2}
- c_{f}} \, kL$, where each KK wavefunction contributes a factor
$\sqrt{2kL}$, the last factor corresponds to the $c_{f}$-dependent
zero-mode wavefunction, and the factor $1/(5kL)$ is a measure of the
region that contributes to the integral in
Eq.~(\ref{wavefunctionEnhancement}).\footnote{The factor of $1/5$ is
determined by comparison to the exact result,
Eq.~(\ref{wavefunctionEnhancement}), and reproduces it within $30\%$
for $-0.5 \lsim c_{f} \lsim 0.4$.} Compared to the annihilation into a
pair of IR localized Higgses, Eq.~(\ref{BulkAnnXSHiggs}), the
annihilation into fermion and KK fermion is ``suppressed'' by order
$(N_{c}/2) (\lambda_{\psi}/\lambda^{\prime})^{2}
(M_{\Phi}/\Lambda)^{2} (1/\Lambda L)^{2}$, where it was assumed that
the annihilation into fermions is not near threshold, and we used our
estimate for $\eta$ and take $c_{f}$ of order one.  The NDA estimate
for $\lambda_{\psi}$ is $24\pi^{3}$, which is the same as for
$\lambda^{\prime}$.  However, the discussion after
Eq.~(\ref{BulkPhiHInteraction}) indicates that a correct thermal relic
abundance requires a much smaller coupling $\lambda^{\prime} \lsim
\Lambda/(2k)$.  Taking $\Lambda \lsim 10k$, $M_{\Phi} \sim \tilde{k}$,
$kL \approx 34$, one can see that the annihilation into an IR
localized fermion and its lightest KK mode could dominate the
annihilation cross section of $\Phi$ particles.  Other operators that
contribute to the self-annihilation cross section are expected to give
subdominant contributions when $M_{\Phi} \ll \tilde{\Lambda}$.
Nevertheless, the operators that lead to annihilation into gauge
bosons can be interesting from the point of view of DM signals, and
are discussed in subsequent sections.

\begin{figure}[t]
\centerline{ 
\includegraphics[width=0.45 \textwidth]{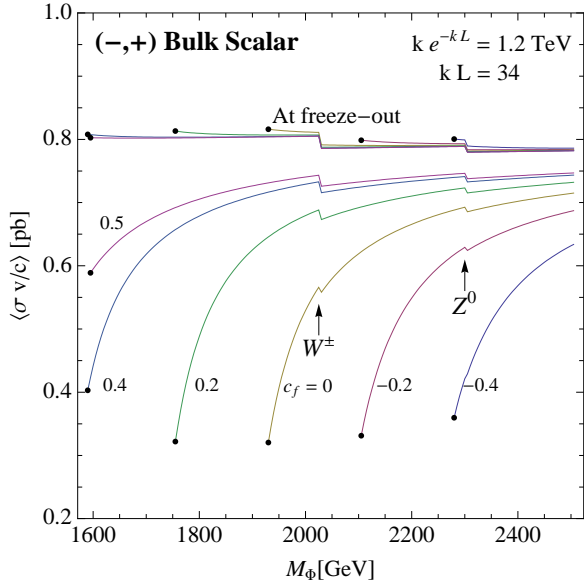}
} 
\caption{Annihilation cross section, $\langle \sigma_{\Phi\Phi
\rightarrow \psi^{(0)} \psi^{(1)}} v\rangle$, for \textit{bulk} DM, as
a function of $M_{\Phi}$, imposing the WMAP constraint on the DM relic
density.  The curves marked as ``freeze-out'' correspond to the
annihilation cross section at the time of freeze-out (where the
typical velocities were of order $v/c \sim \sqrt{2/25} \sim 0.3$).
The lower curves correspond to the annihilation cross section in the
ultra non-relativistic regime, as would be relevant for today's
conditions.  The various curves correspond to different choices of the
fermion localization parameter $c_{f}$ that controls their masses and
couplings.  The arrows indicate the points where the freeze-out
temperature ($\sim M_{\Phi}/25$) crosses the $W^{\pm}$ and $Z^0$
thresholds.  The curves are terminated (with black dots) when
$\lambda_{\psi} = 24\pi^{3}$, which we define as the strong coupling
regime (see text). We assume that $\Lambda = 8 k$.}
\label{fig:bulkscalar}
\end{figure}
We explore here the exotic picture where the $\Phi$s are kept in
thermal equilibrium dominantly by annihilation into a fermion and its
lightest KK mode.  For concreteness, we imagine here scenario 1
discussed above, where the annihilation proceeds mainly into tops and
its lightest KK resonance, but the same would hold in scenario 3 with
one of the lightest leptons (either electrons or neutrinos) replacing
the top.  The results can also be applied in a straightforward way to
scenario 2 with all fermions localized identically: one should just
include a multiplicity factor $3 \times [ 3 \times 4 + 3 ] = 45$.  

In scenario 1, the processes taken into account are $\Phi\Phi
\rightarrow \bar{T}^{(1)}_{L} t_{R}$ and $\Phi\Phi \rightarrow
T^{(1)}_{L} \bar{t}_{R}$, where $T^{(1)}$ is the first KK excitation
of the RH top tower.  The annihilation cross section depends on the
lightest KK scalar and fermion masses $M_{\Phi}$ and $M_{T^{(1)}}$,
which are both of order $\tilde{k}$.  We fix $\tilde{k}$ and obtain
different values of $M_{\Phi}$ as described in
Appendix~\ref{sec:scalars}.  The KK fermion mass has some dependence
on $c_{t}$, which controls the localization of the $t_{R}$
wavefunction.  The overall strength of the cross section depends on
the combination $\lambda_{t}/[\tilde{\Lambda} (\Lambda L)] =
\lambda_{t}/[\tilde{k} (k L)] (k/\Lambda)^{2}$.  Assuming again that
the $\Phi$s account completely for the observed DM energy density,
one can then fix the quantity\footnote{If $\Lambda$ is defined as the
scale where the $SU(3)_{C}$ gauge factor gets strong, then NDA gives
$\Lambda L \sim 24\pi^{3}/(3g^{2}_{s})$, where $g_{s}$ is the 4D color
coupling at the KK scale.  For $kL \approx 34$ this corresponds to
$\Lambda/k \approx 8$.} $\lambda_{t}(k/\Lambda)^{2}$ using the WMAP
result and Eq.~(\ref{Omegah2}) with $\langle \sigma v \rangle = a + 3
b/x_{F}$ where the coefficients $a$ and $b$ are given in
Eq.~(\ref{abFermions}).  In Fig.~\ref{fig:bulkscalar} we show the
result for several values of the fermion localization parameter $c_{f}
= c_{t}$ (for the RH top most likely $c_{t}$ is close to $0$).  As
expected, the annihilation cross section at freeze-out is $\langle
\sigma v/c \rangle \approx 0.8~{\rm pb}$.  However, unlike the case of
annihilation into scalar particles such as the Higgs field discussed
in the previous subsection, both the $a$ and $b$ terms give a
comparable contribution.  As a result, the annihilation cross section
at very low-temperatures, being dominated by the $a$ term, is somewhat
different from the cross section at freeze-out.  This is relevant for
annihilation under today's conditions, and is also shown in
Fig.~\ref{fig:bulkscalar}.  The curves marked as ``At freeze-out''
correspond to the annihilation cross section at the time of $\Phi$
decoupling (when the typical velocities were $v/c \sim \sqrt{2/25}
\sim 0.3$), while the curves in the lower part of the plot correspond
to the annihilation cross section in the ultra non-relativistic
regime, and correspond essentially to the $a$-term in
Eq.~(\ref{abFermions}).  The behavior observed in these curves arises
from the fact that as $\Delta M = 2M_{\Phi} - M_{T^{(1)}}$ approaches
zero, the annihilation cross section vanishes.  Specifically $a \sim
\lambda^{2}_{\psi} (\Delta M)^{2}$ and $b \sim \lambda^{2}_{\psi}
\Delta M$.  Thus, near threshold the $b$ term dominates, and the WMAP
constraint requires the scaling $\lambda^{2}_{\psi} \sim 1/\Delta M$.
This explains why the annihilation cross section at very low
temperatures decreases as $M_{\Phi}$ decreases (for fixed
$M_{T^{(1)}}$), since $a \sim \lambda^{2}_{\psi} (\Delta M)^{2} \sim
1/\lambda_{\psi}^2 \sim \Delta M$.  

We terminate the curves at the point where the coupling
$\lambda_{\psi}$ reaches the strong coupling value given by NDA,
$\lambda_{\psi} \sim 24\pi^{3}$, assuming $\Lambda \sim 8 k$.  As
explained above this happens near the threshold for top-KK top
production.  The different curves are terminated at different points
due to the $c_{t}$ dependence of the KK fermion mass $M_{T^{(1)}}$.
Thus, at strong coupling, $\lambda_{\psi}$ cancels the volume
suppression factor $\Lambda L$ in Eq.~(\ref{XSFermions}) that arises
from the fact that the operator Eq.~(\ref{BulkPhiPsiInteraction}) is
suppressed by two powers of $\Lambda$.  The fact that this channel
then dominates over the Higgs pair production channel, in spite of
arising from an operator of higher dimensionality can then be
understood as due to the strong localization near the IR brane of the
RH top quark, as encoded in the parameter $\eta$ of
Eq.~(\ref{wavefunctionEnhancement}) as well as the different values of
the couplings of the associated operators.  Away from threshold the
coupling $\lambda_{\psi}$ is a factor of $5$-$10$ below the NDA value,
so that the perturbative computation can be trusted.

It is therefore plausible that the annihilation into Higgses plays a
subdominant role in the determination of the DM relic density.  Of
course it is straightforward to take both channels into account when
they give a comparable contribution, but we will not do so here and
turn instead to the possible DM signals of these scenarios.  Note
however that sizable brane-localized kinetic terms (that were not
included in the above analysis) are known to lower the lightest KK
masses significantly~\cite{BraneLocalized}.  Thus, even for $\tilde{k}
= 1.2~{\rm TeV}$ (as is suggested by the EW precision constraints as a
lower bound on $\tilde{k}$, and as assumed in
Fig.~\ref{fig:bulkscalar}) the KK masses can easily be somewhat below
a TeV. The qualitative behavior of the ultra non-relativistic cross
section persists: it is of order $0.8~{\rm pb}$, and decreases by a
factor of about two near the threshold for $\Phi\Phi \rightarrow
T^{(1)} t$ annihilation (assuming that this is the main annihilation
channel and that we are in the perturbative regime).  Thus, in the
following phenomenological analysis, we will allow a large range of KK
masses and analyze the consequences for indirect detection.

\section{Indirect Detection}
\label{sec:Indirect}  

Because the scalar couples to the Higgs, interactions relevant for
direct-detection experiments are in principle possible
\cite{Burgess:2000yq}.  However for the heavy scalars we are talking
about the direct detection rate will be too low so we concentrate on
indirect signals.

In this section we consider such possible signals from the DM
candidates discussed in Section~\ref{sec:relic}.  We will argue that
current experiments may be sufficiently sensitive to detect photons
(or possibly positrons) from dark matter annihilation, most likely
when non-renormalizable operators are present.  We will present our
bounds in terms of constraints on the cutoff scale $\Lambda$ appearing
in these operators, which in the RS context can be understood as being
related to the fundamental gravity scale and more generally represents
a scale of strong interactions.

We first concentrate on the most distinctive signals, $\Phi\Phi
\rightarrow \gamma X$ and $\Phi\Phi \rightarrow e^{+}X$, where the
photon(s) and positron are produced from direct 2-body decays and have
well-defined energies.  We also consider the more exotic decay chain
involving a KK lepton, which generally yields a continuous spectrum
(even before propagation thorugh the interstellar medium) except when
this KK lepton is sufficiently heavy to be produced almost at rest so
that the positrons that result from its decay have a spectral
distribution similar to those of primary positrons.

Subsequently we will consider possibilities from the decay of the
Higgs that would occur as a consequence of the dark matter-Higgs
coupling.

\subsection{Photons}
\label{sec:photons}  

We now consider possible photon signals arising from annihilating dark
matter.  For a continuous photon signal the total flux is obtained by
integrating from some detector-dependent threshold energy up to the DM
mass.  In the scenarios discussed in Section~\ref{sec:relic} the
continuous signal is likely too small to see but we comment on such
decays at the end of this subsection.

We start by discussing the more interesting signal arising from the
direct decays of the (slowly moving) DM particle into photons
proceeding from higher dimension operators, in which case the final
photon is nearly monoenergetic.  Both decays into two photons and a
photon and a $Z$ could in principle contribute.  The photon energy in
the first process is approximately equal to $M_{\Phi}$, while in the
second process it is approximately
$M_{\Phi}(1-M^{2}_{Z}/4M^{2}_{\Phi})$.  For DM in the TeV range, the
energy resolution of ACT's is not enough to resolve the two lines and
they both appear to have energy essentially equal to $M_{\Phi}$.  It
is therefore appropriate to add the two photon signals in the flux.

The annihilation of a SM singlet into photons can proceed via higher
dimension operators which we write as
\beqa
- \frac{e^{2} \kappa}{8\tilde{\Lambda}^{2}} \, \Phi^{2} F_{\mu\nu} F^{\mu\nu}
-\frac{e^{2} \kappa^{\prime} }{4 s_{W} c_{W} \tilde{\Lambda}^{2}} \, 
\Phi^{2} Z_{\mu\nu} F^{\mu\nu}~,
\label{HigherDimPhotons}
\eeqa
where $F_{\mu\nu}$ and $Z_{\mu\nu}$ are the photon and $Z$ gauge boson
field strengths, $s_{W}$ is the sine of the weak mixing angle,
$\tilde{\Lambda}$ is the effective 4D cutoff scale, and $\kappa$,
$\kappa^{\prime}$ are couplings of order one.  The operators in
Eq.~(\ref{HigherDimPhotons}) are the 4D effective operators induced by
bulk or brane-localized operators, depending on whether $\Phi$ arises
from a bulk field or is localized on the IR brane.  In the RS
framework the cutoff might be expected to be around the TeV scale and
not far from the mass of $\Phi$, so that the resulting annihilation
into photons need not be extremely suppressed.

In the ultra non-relativistic limit (DM particle velocities in the
galaxy are of order $v \sim 10^{-3} c$), the interaction terms in
Eq.~(\ref{HigherDimPhotons}) give rise to the cross sections
\beqa
\langle \sigma_{2\gamma} v/c \rangle &\approx& 
\left( \frac{M_{\Phi}}{\tilde{\Lambda}} \right)^{4} 
\frac{3\pi\alpha^{2}\kappa^{2}}{M^{2}_{\Phi}}
\nonumber \\
&\approx&
0.2~{\rm pb} \left( \frac{1~{\rm TeV}}{M_{\Phi}} \right)^{2} 
\left( \frac{M_{\Phi}}{\tilde{\Lambda}} \right)^{4} \kappa^{2}~,
\label{XS-photons}
\eeqa
where $\alpha$ is the fine structure constant, and an annihilation
cross section into $\gamma Z$
\beqa
\langle \sigma_{\gamma Z} v/c \rangle &\approx& 
\left( \frac{M_{\Phi}}{\tilde{\Lambda}} \right)^{4} 
\frac{6\pi\alpha^{2}\kappa^{\prime 2}}{s^{2}_{W} c^{2}_{W} M^{2}_{\Phi}} \, 
\left(1 - \frac{M^{2}_{Z}}{4M^{2}_{\Phi}}\right)
\nonumber \\ [0.5em]
&\approx&
2.5~{\rm pb} \left( \frac{1~{\rm TeV}}{M_{\Phi}} \right)^{2} 
\left( \frac{M_{\Phi}}{\tilde{\Lambda}} \right)^{4} \kappa^{\prime 2}~.
\label{XS-Zphotons}
\eeqa
Notice that besides the enhancement in the $\gamma Z$ channel due to
the gauge coupling (the factor $1/s^2_{W}c^2_{W} \approx 5.6$), there
is an additional factor of 2 difference due to the identical particle
nature of the final state photons in the $2\gamma$ channel.  This
factor is compensated by the explicit factor of 2 in
Eq.~(\ref{PhotonFlux}) that accounts for the two photons in the final
state.

The rates given in Eqs.~(\ref{XS-photons}) and (\ref{XS-Zphotons}) are
small when the $\Phi$ mass is low due to the strong
$(M_{\Phi}/\tilde{\Lambda})^{4}$ dependence.  However, if
$\tilde{\Lambda}$ is not much above $M_{\Phi}$, ground based Cherenkov
detectors can be sensitive to this signal.  We will now interpret
current bounds in terms of the implications for the cutoff scale
$\tilde{\Lambda}$.

\begin{figure}[t]
\centerline{ 
\includegraphics[width=0.48 \textwidth]{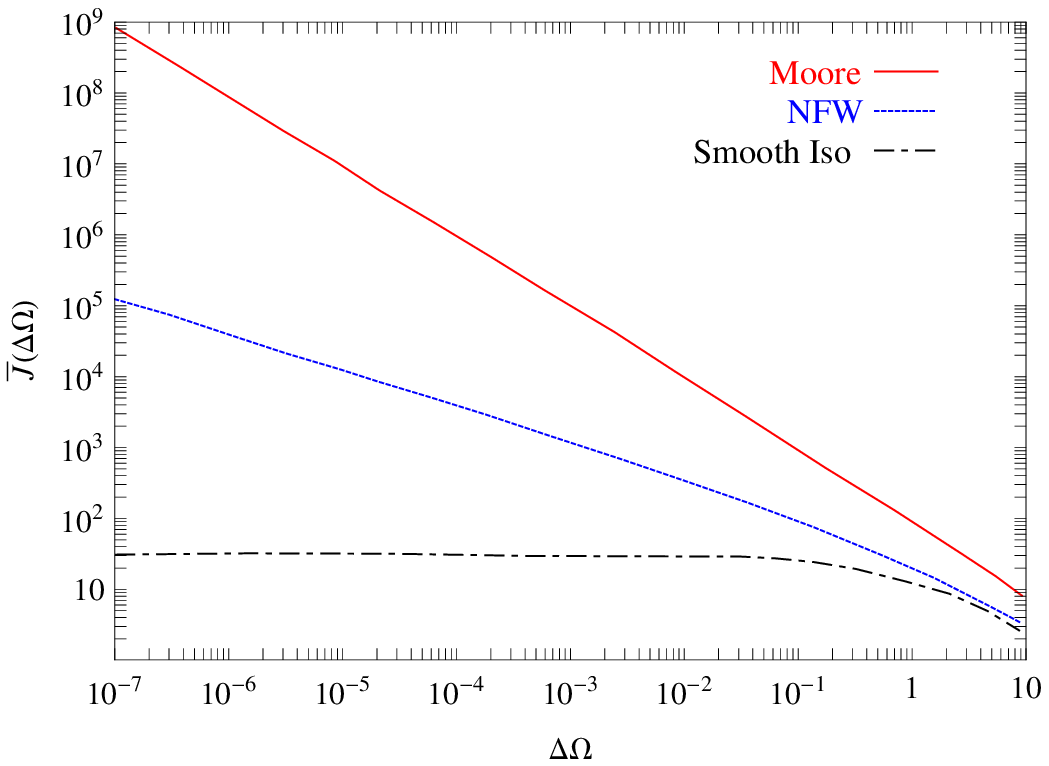}
\hspace*{0.5cm}
\includegraphics[width=0.48 \textwidth]{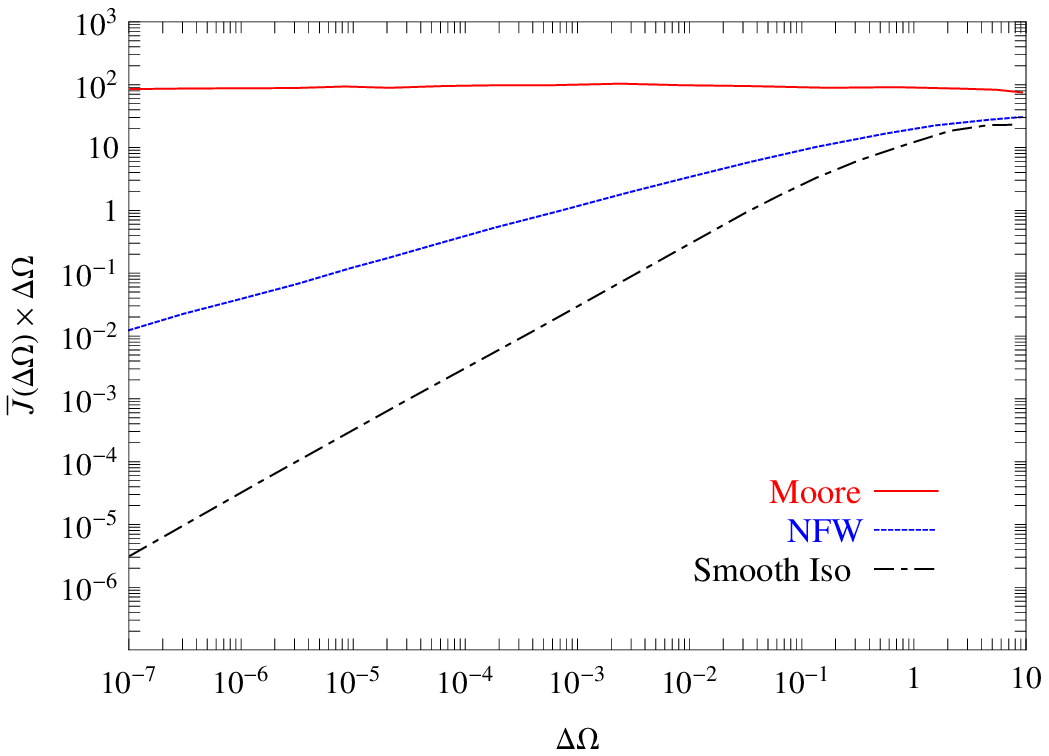}
} 
\caption{Left panel: $\bar{J}(\Delta\Omega)$ as a function of
$\Delta\Omega$ for three different halo profiles (taken from
Ref.~\cite{Thomas:2005te}).  Right panel: $\bar{J}(\Delta\Omega)
\times \Delta \Omega$ as a function of $\Delta\Omega$.}
\label{fig:Jpsi}
\end{figure}

The differential photon flux from a direction that forms an angle
$\psi$ with the galactic plane is
\beqa
\frac{d\Phi_{\gamma}}{d\Omega dE} &=&
\sum_{i} \, \langle \sigma_{i} v \rangle \frac{d N^{i}_{\gamma}}{d E}
\frac{1}{4\pi M^{2}_{\Phi}} \int^{\infty}_{0} \! dl \rho^{2}(r)~,
\eeqa
where $r^{2} = l^{2} + r^{2}_{0} - 2 l r_{0} \cos\psi$, with $r_{0}
\approx 8.5~\rm{kpc}$ the distance from the Earth to the galactic
center.  The integration is along the line of sight, $dl$, and encodes
the information about the DM distribution, assuming a spherical DM
halo of energy density $\rho(r)$.  The particle physics input enters
through the thermally averaged cross section times relative velocity
(for channels labeled by $i$) and the differential photon yield in
channel $i$, $d N^{i}_{\gamma}/d E$ where we add the $\gamma\gamma$ and
$\gamma Z$ signals.  We have
\beqa
\Phi_{\gamma} = 5.66\times 10^{-12}~{\rm cm^{-2} s^{-1}} \left[ 2 
\left( \frac{\langle \sigma_{2\gamma} v/c \rangle}{1~{\rm pb}} \right) 
+ \left( \frac{\langle \sigma_{\gamma Z} v/c \rangle}{1~{\rm pb}} \right) \right]
\left( \frac{1~{\rm TeV}}{M_{\Phi}} \right)^{2} \bar{J}(\Delta\Omega) \Delta\Omega~,
\label{PhotonFlux}
\eeqa
where the factor of $2$ corresponds to the two photons per decay in
the $2\gamma$ annihilation channel, $\bar{J}(\Delta\Omega) \equiv
(1/\Delta\Omega) \int_{\Delta\Omega} J(\psi) d\Omega$ integrates over
the angular acceptance of the detector $\Delta\Omega$, and $J(\psi)$
is conventionally defined as
\beqa
J(\psi) = \frac{1}{8.5~{\rm kpc}} \left( \frac{1}{0.3~{\rm GeV/cm^{3}}} \right)^{2} 
\int^{\infty}_{0}dl \rho^{2}(r)~.
\eeqa
The quantity $\bar{J}(\Delta\Omega)$ depends on the DM halo profile
and can vary over several orders of magnitude depending on the halo model,
when looking towards the galactic center.  It has been computed for
several DM halo models, and as a function of $\Delta\Omega$ in
\cite{Bergstrom:1997fj}.  In the left panel of Fig.~\ref{fig:Jpsi} we
reproduce $\bar{J}(\Delta\Omega)$ as a function of $\Delta\Omega$ for
three different halo profiles: the Moore et.  al.
profile~\cite{Moore:1999gc} (a rather cuspy profile), the widely used
Navarro-Frenk-White (NFW) profile~\cite{Navarro:1995iw}, and a smooth
isothermal profile~\cite{Thomas:2005te}.  For reference we also show
in the right panel the product $\bar{J}(\Delta\Omega) \times
\Delta\Omega$ as a function of $\Delta\Omega$.  The angular acceptance
$\Delta\Omega$ depends on the experimental setup.

Several experiments exist that can search for photons from dark matter
annihilation.  Among the particle community, FERMI (formerly known as
GLAST) has recently received a great deal of attention.  FERMI
is a satellite-based detector with excellent angular coverage (greater
than about 2 sr) and fairly good energy resolution ($< 10 \%$)
\cite{GLASTinfo}.  FERMI should have a flux sensitivity of order
$10^{-10}$ photons ${\rm cm}^{-2}~{\rm s}^{-1}$ for photon energies between
about 20 and 300 GeV (with decreasing sensitivity at lower energies
and no sensitivity at higher energies).

Ground-based detectors, on the other hand, have much smaller angular
coverage and worse energy resolution.  However their flux sensitivity
is similar to that of FERMI at around 50 GeV \cite{GLASTBrochure} and
rapidly overtakes FERMI's sensitivity, depending on angular coverage,
at higher energies.  From the right panel in Fig.~\ref{fig:Jpsi}, and
taking the NFW halo profile, we see that when $\Delta\Omega =
10^{-5}~{\rm sr}$, a typical value used in HESS, one has
$\bar{J}(\Delta\Omega) \times \Delta\Omega \approx 10^{-1}$, while for
FERMI with $\Delta\Omega = 2$ one has $\bar{J}(\Delta\Omega) \times
\Delta\Omega \approx 20$.  Therefore, if HESS achieves only $\Delta
\Omega=10^{-5}$, it can be more sensitive than FERMI to the photon
signal we discuss at energies of about 250 GeV~\cite{GLASTBrochure}
(near the end of FERMI's sensitivity range).  Ground-based detection
will be relatively more sensitive with a Moore profile and less so
with an isothermal profile.

HESS and VERITAS could reach a larger angular coverage, since their
fields of view ($5^\circ$ for HESS and $3.5^\circ$ for VERITAS)
correspond to $\Delta \Omega \sim 10^{-2}~{\rm sr}$.  If such angular
acceptances are reached, and assuming an NFW profile so that
$\bar{J}(\Delta\Omega) \times \Delta\Omega \approx 3$ (see
Fig.~\ref{fig:Jpsi}), these ground-based detectors could overtake
FERMI's sensitivity even at photon energies of about 100 GeV. Clearly
the flux sensitivity is better for either of the ground based
experiments\footnote{VERITAS does not always point toward the galactic
center, however, so the flux sensitivy in that regime is not
guranteed} for reasonable dark matter masses above about 100 GeV and
the determining factor of which is better is likely to be the angular
resolution.\footnote{The sensitivity of ACTs to the photon signal at
large $\Delta \Omega$ could be limited by the cosmic ray background,
since subtracting the signal from a nearby region can have a
significant effect for shallow profiles~\cite{Mack:2008wu}.  However,
for more peaked profiles such as NFW, this is expected to be at most
an order one effect.}

Of course, without knowing the dark matter profile, it makes sense to
search in both satellite and ground-based experiments at low energies.
However, it should be borne in mind that most dark matter models
predict a monochromatic photon signal only at one loop so indirect
detection is unlikely to be sufficiently sensitive to this type of
signature of standard thermal dark matter.  Supersymmetric dark matter
annihilation into photons offers perhaps the best possible
loop-suppressed scenario because the loop can be
enhanced~\cite{Bergstrom:1997fh,Bern:1997ng} due to a reasonably large
numerical factor and because for a higgsino dominated neutralino an
enhancement in the loop diagram due to near degeneracy with an
intermediate state can lead to a cross section that saturates with
$1/m_W^2$ dependence (rather than suppression by the potentially
bigger dark matter mass).  This signal is potentially observable,
however, only for light dark matter candidates for which the flux is
big (and where FERMI is sensitive).  Otherwise the cross section is
too small.

We note that the direct signal we discuss is at higher energies since
we assume a heavy dark matter candidate and therefore concentrate on
ground-based experiments since they have better sensitivity.  Although
the number density of heavy dark matter particles is lower than that
for lighter dark matter candidates, our prediction is a tree-level
effect, albeit through a higher-dimension operator, and the cross
section can be bigger than typical supersymmetric annihilation cross
sections~\cite{Bergstrom:1997fh}, which saturate at about
$10^{-28}~\rm{cm}^{-3}~\rm{s}^{-1}$.  For example, in the first row of
Table~\ref{ACTBenchmarks} we see that for $\tilde{\Lambda} = 2~{\rm
TeV}$ and $M_{\Phi} = 1~{\rm TeV}$ the annihilation cross section is
$5.4 \times 10^{-27}~{\rm cm}^3~{\rm s}^{-1}$.

For instance, for $1~{\rm TeV}$ photons, HESS has a flux sensitivity of about
$10^{-13}~{\rm cm^{-2}~s^{-1}}$.  Using $\Delta\Omega =
10^{-5}~{\rm sr}$ and taking the NFW halo profile, we see from the
right panel in Fig.~\ref{fig:Jpsi} that $\bar{J}(\Delta\Omega) \times
\Delta\Omega \approx 10^{-1}$.  The expected flux is then
$\Phi_{\gamma} = 1.6\times 10^{-12}~{\rm cm^{-2}~s^{-1}} (1~{\rm
TeV}/\tilde{\Lambda})^{4}$, which could be translated into a bound
$\tilde{\Lambda} \gsim 2~{\rm TeV}$ (we also assumed
$\kappa=\kappa^{\prime} = 1$).\footnote{For $M_{\Phi} = 1~{\rm TeV}$
and $\tilde{\Lambda} = 2~{\rm TeV}$, the non-relativistic annihilation
cross section into $\gamma Z$ is $\langle \sigma_{\gamma Z} v/c
\rangle \approx 0.15~{\rm pb}$, which is smaller than the cross
section necessary for $M_{\Phi}$ to account for the observed DM energy
density.  The annihilation into two photons is smaller by a factor of
about ten.  The largest contribution to the annihilation cross section
would come from either annihilation into Higgses or fermion-KK fermion
pairs as discussed in Section~\ref{sec:relic}, thus justifying the
relic density computation discussed there.} Under the same
assumptions, for $2.3~{\rm TeV}$ photons HESS would put a bound
$\tilde{\Lambda} \gsim M_{\Phi} \sim 2.3~{\rm TeV}$.  On the other
hand, the sensitivity could in principle be bigger or smaller
according to the dark matter profile.  For example, for the rather
peaked Moore et.  al.  profile, one has $\bar{J}(\Delta\Omega) \times
\Delta\Omega \approx 10^{2}$ and the non-observation of a line at
$1~{\rm TeV}$ by HESS would correspond to a bound $\tilde{\Lambda}
\gsim 11.3~{\rm TeV}$.  This is the expected scale for
$\tilde{\Lambda}$ in several well-motivated scenarios that take into
account the EW
constraints~\cite{Agashe:2003zs,Agashe:2004rs,Agashe:2006at,Carena:2006bn}.

\begin{table}[t]
\begin{center}
\begin{tabular}{|c|c|c|c|}
\hline
\rule{0mm}{5mm} $\Delta\Omega$ & $\bar{J}(\Delta\Omega) \times \Delta\Omega$ & 
$\tilde{\Lambda}~[{\rm TeV}]$ at $M_{\Phi} = 1~{\rm TeV}$
& $2\langle \sigma_{2\gamma} v \rangle + \langle \sigma_{\gamma Z} v \rangle~[{\rm cm}^{3}~{\rm s}^{-1}]$ \\ [0.2em]
\hline
\rule{0mm}{5mm} $10^{-5}$ & $10^{-1}$ (NFW) &  2 & $5.4 \times 10^{-27}$ $(1.8 \times 10^{-1}~{\rm pb})$ \\ 
\rule{0mm}{5mm} $10^{-3}$ & $1$ (NFW) &  3.5 & $5.8 \times 10^{-28}$ $(1.9 \times 10^{-2}~{\rm pb})$ \\ 
\rule{0mm}{5mm} any & $10^{2}$ (Moore) &  11.3 & $5.3 \times 10^{-30}$ $(1.8 \times 10^{-4}~{\rm pb})$ \\ [0.2em]
\hline
\end{tabular}
\end{center}
\caption{Sensitivity of HESS or VERITAS to the cutoff scale
$\tilde{\Lambda}$ for representative $\Delta\Omega$'s (NFW and Moore
et.  al.  halo profiles).  A DM candidate with mass $M_{\Phi} = 1~{\rm
TeV}$ annihilating into monoenergetic $1~{\rm TeV}$ photons is
assumed.  We assume $\kappa = \kappa^{\prime} = 1$ (see text).  The
last column gives the thermally averaged annihilation cross section
into photons for the corresponding $\tilde{\Lambda}$ (and for
$M_{\Phi} = 1~{\rm TeV}$).}
\label{ACTBenchmarks}
\end{table}
For HESS or VERITAS operating at $\Delta\Omega = 10^{-3}~{\rm sr}$ and
using again the NFW halo model with $\bar{J}(\Delta\Omega) \times
\Delta\Omega \approx 1$, the expected flux would be $\Phi_{\gamma} =
1.6\times 10^{-11}~{\rm cm^{-2}~s^{-1}} (1~{\rm TeV}/M_{\Phi})^{2}
\times (M_{\Phi}/\tilde{\Lambda})^{4}$.  For $1~{\rm TeV}$ photons,
HESS or VERITAS would be sensitive to $\tilde{\Lambda} \sim 3.5~{\rm
TeV}$.  We summarize these observations in Table~\ref{ACTBenchmarks}.

Ground-based Cherenkov detectors capable of operating at larger
$\Delta\Omega$ can start probing theoretically interesting values of
$\tilde{\Lambda}$ even for halo profiles not as peaked as the Moore
et.  al.  profile.  From Fig.~\ref{fig:Jpsi} we see that for $\Delta
\Omega \sim 1$, several halo profiles converge to
$\bar{J}(\Delta\Omega) \times \Delta\Omega \approx 10$.  An additional
factor of $6$ improvement in the flux sensitivity would then make
scales $\tilde{\Lambda} \sim 10~{\rm TeV}$ accessible.  Of course,
larger $\Delta\Omega$ means also larger background, but hopefully the
very characteristic line signal can be extracted if there are enough
events (see Ref.~\cite{Thomas:2005te}).

We finally mention the possibility of observing photons from Higgs
decays (assuming that the main channel for DM annihilation is into
Higgses, as in Subsection~\ref{sec:BraneFields}, so that $\langle
\sigma_{HH} v \rangle \approx 0.8~{\rm pb}$).  For instance, for a
SM-like Higgs with mass around $m_{h} = 135~{\rm GeV}$, the branching
fractions into $\gamma\gamma$ or $Z\gamma$ are of order $10^{-3}$
each.  The photons from these channels present a flat spectrum between
$E^{\gamma\gamma}_{\rm min} = \frac{1}{2} M_{\Phi} (1-\beta)$ and
$E^{\gamma\gamma}_{\rm max} = \frac{1}{2} M_{\Phi} (1+\beta)$ for the
$\gamma\gamma$ signal, or between $E^{Z\gamma}_{\rm min} = \frac{1}{2}
M_{\Phi} (1-m^{2}_{Z}/m^{2}_{H}) (1-\beta)$ and $E^{Z\gamma}_{\rm max}
= \frac{1}{2} M_{\Phi} (1-m^{2}_{Z}/m^{2}_{H}) (1+\beta)$ for the
$Z\gamma$ signal.  Here $\beta = \sqrt{1-m^{2}_{H}/m^{2}_{\Phi}}$ is
the velocity of the Higgs in the DM rest frame.  

\begin{figure}[t]
\centerline{ 
\includegraphics[width=0.48 \textwidth]{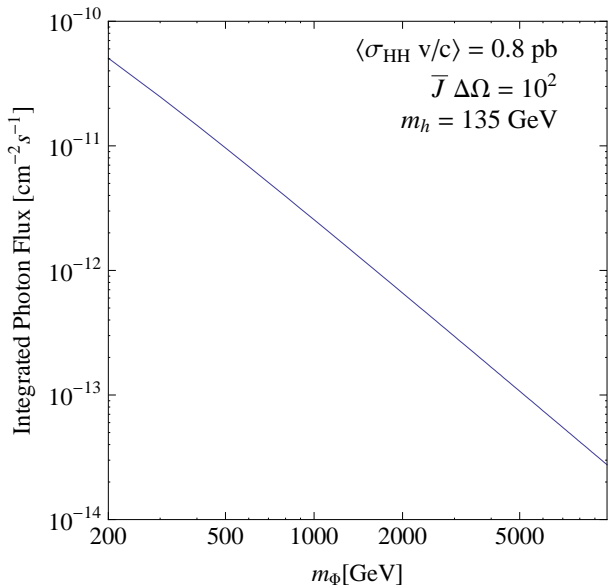}
} 
\caption{Integrated photon flux from $\Phi\Phi \rightarrow HH$ and $H
\rightarrow \gamma\gamma$ or $H \rightarrow Z\gamma$, as a function of
$M_{\Phi}$.  We assume that the branching fractions for these decay
modes are $10^{-3}$ and take $\bar{J}(\Delta\Omega)\times \Delta\Omega
= 10^{2}$.}
\label{fig:HiggsPhotons}
\end{figure}

We show in Fig.~\ref{fig:HiggsPhotons} the total flux integrated from
a threshold energy $E_{\rm th} = 50~{\rm GeV}$ up to
$E^{\gamma\gamma}_{\rm max}$, as a function of $M_{\Phi}$.  Here we
optimistically assume $\bar{J}(\Delta\Omega) \times \Delta\Omega =
10^{2}$ as would be appropriate for the Moore profile though of course
with other profiles the signal would be smaller.  ACTs such as HESS or
VERITAS would be sensitive to such a signal, but if the halo profile
is less peaked or if the Higgs branching fraction into photons is
smaller, this continuous signal becomes challenging.  Nonetheless
since this is a generic prediction of this type of model that doesn't
rely on higher-dimension operators exploring the possibility of
detecting such a signal is extremely worthwhile.
%

\subsection{Positrons}
\label{sec:positrons}  

Recently there has been intriguing evidence for an excess positron
signal at energies up to about 80 GeV \cite{Adriani:2008zr}.  Clearly
it is of interest to determine whether such positrons can arise from
dark matter annihilation.  We do not anticipate that heavy dark matter
particles will explain this excess, since a positron signal, if it
exists, will be concentrated at higher energies.

Nonetheless it is of interest to explore this positron signal to see
how it compares to background and to see whether in principle the
signal could be detectable at high energies.  We also briefly consider
lighter particles (with less theoretical motivation in our context of
strongly intereacting TeV scale physics) with similar interactions to
KK dark matter particles and find that even without a big boost factor
annihilations of $\sim 100$ GeV dark matter matches the PAMELA data.

With this in mind, we note that for the bulk scalar fields discussed in
Subsection~\ref{sec:BulkFields}, other operators involving SM fields
can be interesting from the point of view of DM signals, besides the
operators leading to direct annihilation of the $\Phi$ particles into
photons discussed in the previous subsection.  Specifically, the
higher-dimension operators of the type
Eq.~(\ref{BulkPhiKKPsiInteraction}), coupling a pair of $\Phi$s to an
electron and its lowest KK mode, can lead to an interesting positron
signal.  The dominant annihilation channel involves the fermions
closest to the IR brane (we discussed in
Subsection~\ref{sec:BulkFields} how the DM relic density can be
determined by the annihilation into a SM fermion and one of its KK
modes).

In Subsection~\ref{sec:BulkFields} we defined three scenarios
that differ on how the fermions are localized in the extra dimension.
Of these, the most favorable one to obtain a sizable positron signal
from DM annihilation is scenario 3.  But we will see that only the
electrons need to be somewhat localized near the IR brane for the
positron signal to be interesting, and this can occur in scenario 2 as
well.\footnote{A hybrid case (of scenarios 1 and 2) with one of the
lepton chiralities localized somewhat near the IR brane and the
opposite chirality localized near the UV brane (to generate the small
lepton masses by the exponential wavefunction suppression) also falls
in this category.  For instance if $c_{l_{R}} \approx 0.4$ while
$c_{t_{R}} \approx 0$, we have $\eta_{t_{R}}/\eta_{l_{R}} \approx
[(1-2c_{t_{R}})/(1-2c_{l_{R}})]^{1/2} \sim 2$, where the $\eta$
parameters were defined in Eq.~(\ref{wavefunctionEnhancement}).
Therefore, the annihilation into positrons can plausibly be suppressed
compared to the dominant top channel by only a factor $\sim 4 N_{c}
\sim 10$, if the unknown dimensionless coefficients $\lambda_{e}$ and
$\lambda_{t}$ in Eq.~(\ref{BulkPhiPsiInteraction}) are assumed to be
comparable.

If only the RH top and the RH leptons are localized near the IR brane
and one neglects other annihilation channels, the thermal relic
density computation implies $\langle \sigma_{e^{(1)}e} v/c \rangle =
\langle \sigma_{\mu^{(1)}\mu} v/c \rangle = \langle \sigma_{\tau^{(1)}
\tau} v/c \rangle \approx 0.06~{\rm pb}$ and $ \langle \sigma_{t^{(1)}
t} v/c \rangle \approx 0.6~{\rm pb}$.} Note also that the positron
signal is sensitive to the \textit{local} DM distribution (and not
very much to how peaked the DM halo is at the galactic center).  It is
common to parameterize the effects of DM inhomogeneities by an
(energy-independent) ``boost'' factor $B = \langle \rho^{2} \rangle/
\langle \rho \rangle^{2}$.  Studies of such enhancements for gamma
rays indicate that the boost factor might be as large as order
10~\cite{Diemand:2006ik}.  Therefore, the positron signal can receive
an enhancement compared to the case of a smooth DM density
distribution, though the likely size of this enhancement is not
expected to be very large.

With this understanding we proceed to estimate the signal from direct
annihilation into a positron and a KK mode.  The produced positron has
a well-defined energy that depends on the DM and KK fermion masses,
$M_{\Phi}$ and $M_{e^{(1)}}$: $E^{\rm prim.}_{e^{+}} = (4M^{2}_{\Phi}
- M^{2}_{e^{(1)}})/4M_{\Phi}$.  Besides these primary monoenergetic
positrons, we also consider the secondary positrons arising from the
annihilation of $\Phi$ particles into an electron (neutrino) and a
positron KK mode followed by the decay of the associated KK lepton
into a positron and a $Z$ ($W$) gauge boson.\footnote{The decays of
the KK lepton into Higgs are suppressed by the electron Yukawa
coupling.  For gauge KK masses of order $3~{\rm TeV}$, the main decay
channels of the KK lepton involve $Z$ or $W$ (through EWSB mixing of
the $Z/W$ with its KK modes, as opposed to mixing of the lepton and
its KK modes).  When the lepton is an $SU(2)$ doublet we have
$\Gamma(e^{(1)} \rightarrow Z e)/\Gamma(\nu^{(1)} \rightarrow W e)
\approx (m^{4}_{Z}/m^{4}_{W})(T^{3} - s^{2}_{W} Q)^{2}/c^{2}_{W}$,
leading to ${\rm BR}(\nu^{(1)} \rightarrow W e) \approx 75\%$ and
${\rm BR}(e^{(1)} \rightarrow Z e) \approx 25\%$.  Similarly, the
$SU(2)$ singlet KK positron decays dominantly into $Ze^{+}$.} When the
KK lepton $l^{(1)} = e^{(1)}$ or $\nu^{(1)}$ has a mass slightly below
$2M_{\Phi}$ (about the threshold for DM annihilation into lepton and
KK lepton), it is produced nearly at rest and the resulting positron
from its decay has a relatively well-defined energy.  In detail, the
energy of the KK lepton produced in DM annihilation is $E_{l^{(1)}} =
(4M^{2}_{\Phi} + M^{2}_{l^{(1)}})/4M_{\Phi}$, while its momentum is $p
= (4M^{2}_{\Phi} - M^{2}_{l^{(1)}})/4M_{\Phi}$.  For the two-body
decays $e^{(1)} \rightarrow Z e$ or $\nu^{(1)} \rightarrow W e$, one
finds the typical flat spectral distribution
\beqa
f_{2}(E_{0}) = 
\left\{
\begin{array}{cl}
(E_{l^{(1)}}\beta_{l^{(1)}})^{-1}  &  E_{-} \leq E_{0} \leq E_{+}  \\ [0.5em]
0  &  {\rm otherwise}
\end{array}
\right.~,
\label{f2}
\eeqa
where $E_{0}$ is the positron energy.  Neglecting the masses of the
decay products, the maximum positron energy is $E_{+} = \frac{1}{2}
E_{l^{(1)}} (1 + \beta_{l^{(1)}}) = M_{\Phi}$, while the minimum
positron energy is $E_{-} = \frac{1}{2} E_{l^{(1)}} (1 -
\beta_{l^{(1)}}) = M^{2}_{l^{(1)}}/4M_{\Phi}$.  Here $\beta_{l^{(1)}}
= p/E_{l^{(1)}} = (4M^{2}_{\Phi} - M^{2}_{l^{(1)}})/(4M^{2}_{\Phi} +
M^{2}_{l^{(1)}})$ is the velocity of the KK lepton (in the rest frame
of $\Phi$).  Notice that the upper endpoint is determined by the DM
mass only, and that in the limit $M_{l^{(1)}} \rightarrow 2M_{\Phi}$
one has $f_{2}(E_{0}) \rightarrow \delta(E_{0} - M_{\Phi})$.  Further
decays of the $W$s and $Z$s lead to additional positrons that have a
softer spectrum and give a subdominant contribution due to the small
branching fractions involved.  We do not include positrons from $W$ or
$Z$ decay in the following analysis.  Note also that primary and
secondary electrons with the exact same characteristics as the
positrons above are also produced.

The positron energy is distorted as it propagates through the
interstellar medium before detection.  In general, for an initial
spectral distribution $f_{i}(E_{0})$, normalized according to
$\int^{\infty}_{0} dE_{0} f_{i}(E_{0}) = 1$, the differential positron
flux at the solar position is obtained from
\beqa
\frac{d\Phi_{e^+}}{d\Omega dE} = \frac{B \rho^2_{0}}{m^{2}_{\Phi}}
\sum_{i} \langle \sigma_{i} v\rangle B^{i}_{e^{+}} 
\int dE_{0} f_{i}(E_{0}) G(E_{0}, E)~,
\eeqa
where $\rho_{0}$ is the average DM mass density, $B$ is the boost
factor, $\langle \sigma_{i} v\rangle$ is the $i$-th channel thermally
averaged (ultra non-relativistic) DM annihilation cross section times
relative velocity, $B^{i}_{e^{+}}$ is the corresponding branching
fraction into positrons, and $G(E_{0}, E)$ is a Green function that
includes the details of the DM mass distribution in the galactic halo,
takes into account the propagation of the positrons through the
interstellar medium in the galaxy, and describes how their energy $E$
is shifted under diffusion, various spatially and energy-dependent
energy loss mechanisms, reacceleration, etc.  The direct annihilation
into positrons plus their lightest KK mode simply corresponds to
$f(E_{0}) = \delta(E_{0} - E^{\rm prim.}_{e^{+}})$, while secondary
positrons arising from the decay of the KK lepton are described by
Eq.~(\ref{f2}).

In Ref.~\cite{Moskalenko:1999sb}, Moskalenko and Strong modeled the
propagation of positrons through the interstellar medium for several
galactic halo DM mass distributions.  They provided a simple
parameterization for the Green function that reproduces the more
detailed simulation\footnote{The code used in the simulation aims at
reproducing simultaneously observational data related to cosmic ray
origins and propagation such as: direct measurements of nuclei,
antiprotons, electrons and positrons, as well as indirect measurements
via $\gamma$ rays and synchrotron radiation.} to within 10\%:
\beqa
10^{-25} E^2 G(E_{0},E) = 10^{a (\ln E)^{2} + b \ln E + c} \, 
\theta(E-E_{0}) + 10^{w (\ln E)^{2} + x \ln E + y} \, \theta(E_{0} - E)~,
\eeqa
where $G(E_{0},E)$ is given in units of ${\rm cm~sr^{-1} GeV^{-1}}$,
$E$ is the local positron energy in GeV, and the coefficients $a$,
$b$, $c$, $w$, $x$ and $y$ are functions of $E_{0}$ (the initial
positron energy) that are tabulated in Tables II and III of
Ref.~\cite{Moskalenko:1999sb}.  For definiteness, we consider the
``isothermal'' model, which is characterized by a spherically
symmetric DM mass distribution given by
\beqa
\rho(r) = \rho_{0} \, \frac{r^{2}_{c} + R^{2}_{\odot}}{r^{2}_{c} + r^{2}}~,
\eeqa
where $r_{c}$ is the core radius and $R_{\odot} = 8.5~{\rm kpc}$ is
the solar distance to the galactic center (the parameters $r_{c}$ and
$\rho_{0}$ are obtained by fitting to the rotation curve, and for the
isothermal model $r_{c} = 2.8~{\rm kpc}$ and $\rho_{0} = 0.43~{\rm
GeV~cm^{-3}}$).  We use a galactic halo size of $z_{h} = 10~{\rm
kpc}$, which is on the upper limit of the $4-10~{\rm kpc}$ range
favored by the analysis in \cite{Moskalenko:1999sb}.

The local positron flux then takes the form
\beqa
E^{2} \frac{d\Phi_{e^+}}{d\Omega dE} = 2.7\times 10^{-8} 
\left( \frac{\rho_{0}}{0.3~{\rm GeV/cm^{3}}} \right)^2 
\left(\frac{1~{\rm TeV}}{m_{\Phi}} \right)^{2}
\sum_{i} \left(\frac{B \langle \sigma_{i} v/c\rangle}{1~{\rm pb}} \right) 
B^{i}_{e^{+}} F_{i}(E)~,
\label{PositronFlux}
\eeqa
where the units on the r.h.s. are ${\rm GeV~cm^{-2} s^{-1} sr^{-1}}$,
and the dimensionless $F_{i}(E)$ is defined by
\beqa
F_{i}(E) = 10^{-25} E^{2} \int dE_{0} f_{i}(E_{0}) G(E_{0}, E)~.
\label{positronPropagation}
\eeqa

Experimental observations are commonly presented in the form of the
positron fraction, $e^+/(e^{-} + e^{+})$, where the electron and
positron fluxes include both background and signal.  This quantity has
the advantage that systematic uncertainties cancel out.  For the electron 
and positron background spectral distributions we use the simple 
parameterizations given in
\cite{Baltz:1998xv}:
\beqa
\left( \frac{d\Phi_{e^-}}{d\Omega dE} \right)_{\textrm{prim. bkg }} &=& 
\frac{0.16 \, E^{-1.1}}{1 + 11 \, E^{0.9} + 3.2 \, E^{2.15}}~,
\nonumber \\
\left( \frac{d\Phi_{e^-}}{d\Omega dE} \right)_{\textrm{sec. bkg }} &=& 
\frac{0.70 \, E^{0.7}}{1 + 110 \, E^{1.5} + 600 \, E^{2.9} + 580 \, E^{4.2}}~,
\nonumber \\
\left( \frac{d\Phi_{e^+}}{d\Omega dE} \right)_{\textrm{sec. bkg }} &=& 
\frac{4.5 \, E^{0.7}}{1 + 650 \, E^{2.3} + 1500 \, E^{4.2}}~,
\nonumber
\eeqa
where $E$ is in GeV and the units of the l.h.s are ${\rm
GeV^{-1}~cm^{-2}~s^{-1}~sr^{-1}}$.

\begin{figure}[t]
\centerline{ 
\includegraphics[width=0.47 \textwidth]{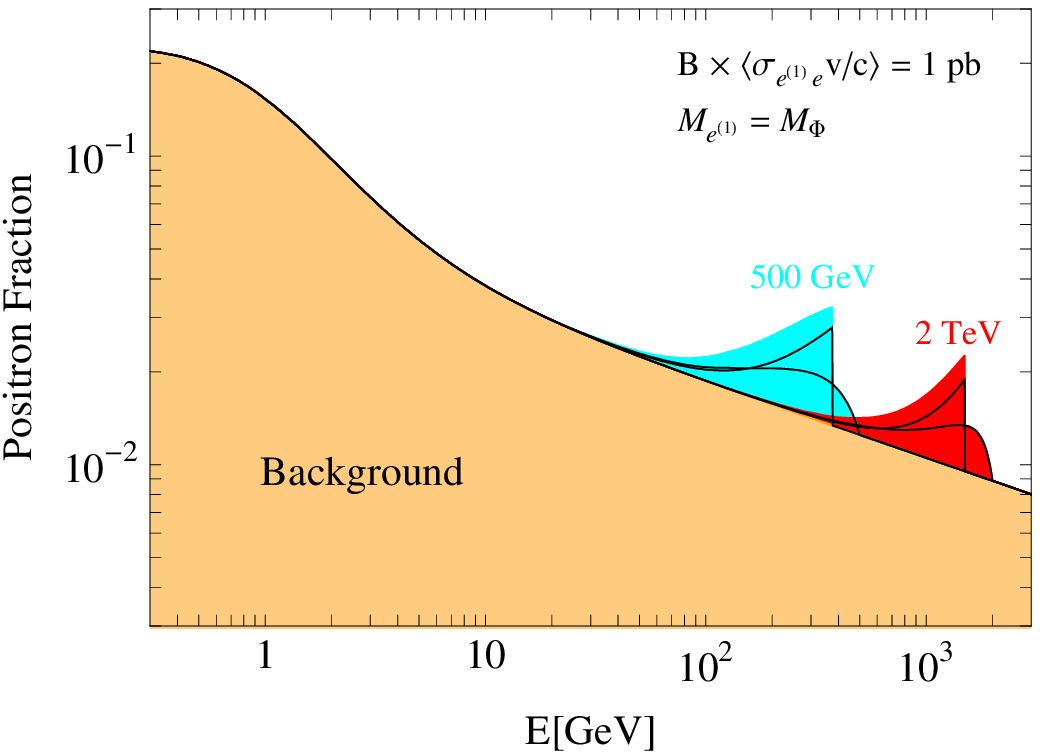}
\hspace*{0.5cm}
\includegraphics[width=0.47 \textwidth]{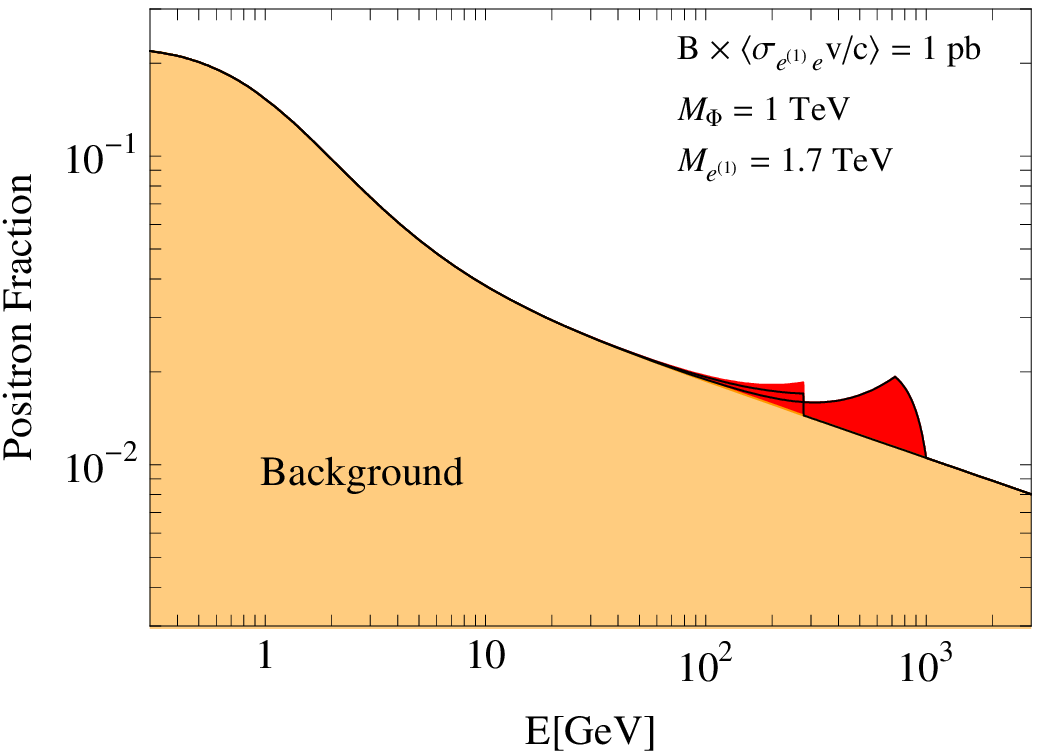}
} 
\caption{Left panel: positron fraction including the primary
positrons/electrons from the annihilation $\Phi\Phi \rightarrow
e^{\pm} e^{(1)}$ and the secondary positrons/electrons from the
annihilation $\Phi\Phi \rightarrow l^{(1)} l^{(0)}$, followed by a
two-body decay $l^{(1)} \rightarrow e^{\pm} X$.  We show the spectra
for two DM masses, $M_{\Phi} = 500~{\rm GeV}$ and $M_{\Phi} = 2~{\rm
TeV}$, assuming that $M_{e^{(1)}} = M_{\Phi}$ and a boost factor such
that $B \times \langle \sigma_{e^{(1)}e} v/c \rangle = 1~{\rm
pb}$, with $\rho_{0} = 0.3~{\rm GeV/cm^{3}}$.  The solid lines
represent the individual contributions from primary and secondary
production.  Right panel: positron fraction for $M_{\Phi} = 1~{\rm
TeV}$ and $M_{l^{(1)}} = 0.85\times(2M_{\Phi}) = 1.7~{\rm TeV}$ showing
clear peaks at $E^{\rm prim.}_{e^{+}} = (4M^{2}_{\Phi} -
M^{2}_{e^{(1)}})/4M_{\Phi}$ and near $M_{\Phi}$ (the endpoint is
exactly at $M_{\Phi}$).}
\label{fig:positronSignal}
\end{figure}
In the following we envision a scenario in which the dominant DM
annihilation channel is into pairs of electron/positron plus a KK
mode, as could be expected in scenario 3 defined in
Subsection~\ref{sec:BulkFields}.  In this case, the WMAP relic
abundance requires $\langle \sigma_{e^{(1)}e} v/c \rangle \approx
1~{\rm pb}$.  Somewhat more generally, the results are valid for $B
\times \langle \sigma_{e^{(1)}e} v/c \rangle = 1~{\rm pb}$, where $B$
is the boost factor.  Since $B$ is expected to be order a few, our
results can illustrate situations where the electron/positron channel
is one among a few dominant annihilation channels (e.g. if the other
lepton channels are equally important).

In the left panel of Fig.~\ref{fig:positronSignal} we show the
expected positron fraction signal, assuming DM masses $M_{\Phi} =
500~{\rm GeV}$ and $M_{\Phi} = 2~{\rm TeV}$.  We also include the
secondary positrons arising from the decay of the KK lepton, assuming
$M_{e^{(1)}} = M_{\Phi}$.  Interestingly, there is a rather clear peak
above background that could be observable in the sub TeV range.  A
moderate boost factor would make such a feature even more prominent.
In the right panel of Fig.~\ref{fig:positronSignal} we show another
example where $M_{e^{(1)}} = 0.85 \times 2M_{\Phi}$ is closer to
threshold.  This case exhibits more clearly the two peaks discussed
above, one at $E^{\rm prim.}_{e^{+}}$ from the primary positrons and
another near $M_{\Phi}$ for the secondary positrons from the KK lepton
decay.  As remarked above, the upper endpoint gives a direct
measurement of $M_{\Phi}$.  The first peak then gives information
about the KK lepton mass.  With a handle on the DM and KK fermion
masses it would be possible to get information about the effective
cutoff scale $(\lambda_{e}/\Lambda L)^{-1} \tilde{\Lambda}$ [see
coefficient $a$ in Eq.~(\ref{abFermions})], modulo the uncertainty
associated with the local DM energy density (since these high-energy
positrons come from distances of at most a few kpc, the dependence on
the DM halo model is expected to be milder).

\begin{figure}[t]
\centerline{ 
\includegraphics[width=0.47 \textwidth]{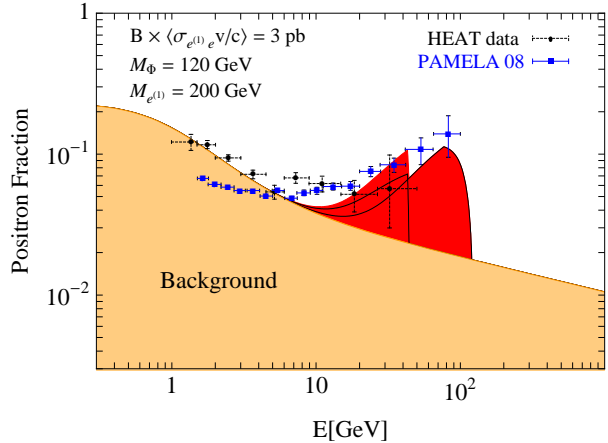}
} 
\caption{Positron fraction due to positrons from a 120 GeV DM
particle annihilating into $e e^{(1)}$, where $e^{(1)}$ is a heavy
electron of mass $M_{e^{(1)}} = 200~{\rm GeV}$.  The lower energy peak
arises from direct annihilation into a positron (plus heavy lepton),
while the higher-energy peak arises from the direct decay of the heavy
lepton into a positron (plus a gauge boson).  Further positrons from
the $W$'s or $Z$'s produced in the heavy lepton decay are not included
and would add a softer contribution.  We assume $B \times \langle
\sigma_{e^{(1)}e} v/c \rangle = 3~{\rm pb}$.}
\label{fig:positronexcess}
\end{figure}

There seems to be evidence in several experiments for an excess in the
positron flux in the tens of GeV energy range
(HEAT~\cite{Barwick:1997ig}, AMS-01~\cite{Aguilar:2007yf}), with the
PAMELA satellite experiment supporting this excess up to energies of
about 80~GeV \cite{Adriani:2008zr}.  As has been emphasized
recently~\cite{Bergstrom:2008gr,Cirelli:2008jk,Barger:2008su} the
observed fluxes are larger than what would be expected from thermal
WIMPs when these lead to positrons mainly through the decays of their
annihilation products (e.g. $W$'s).\footnote{However, the presence of
a relatively long range force among the DM particles can lead to an
enhancement of several orders of magnitude in the annihilation cross
section at very low velocities, which can account for these
observations in certain dark matter
models~\cite{Cirelli:2008jk,Cirelli:2008pk,ArkaniHamed:2008qn}.  Our
DM candidate, having only non-renormalizable interactions, does not
fall into this category.}

Nonetheless, it is interesting to note that a DM candidate with a mass
of about 100 GeV annihilating primarily into electrons/positrons can
explain the observed positron excess with a boost factor of order
unity.  Although such low masses are not expected in our scenario, we
show in Fig.~\ref{fig:positronexcess} the positron signal from the
annihilation of a $120~{\rm GeV}$ DM candidate into a positron and a
``heavy vector-like electron'' of mass $M_{e^{(1)}} = 200~{\rm GeV}$.
We show the HEAT data and the recently released PAMELA data, which
shows a clear increase with energy of the positron fraction up to
energies of at least 80~GeV~\cite{Adriani:2008zr}.  However, the
ATIC-2 balloon experiment~\cite{ATIC} also indicates an excess in the
total electron plus positron flux extending up to energies of about a
TeV, which would not be explained by the self-annihilation into
electrons/positrons of such a light DM candidate.  Antiprotons
produced in decays of the heavy electron (via W gauge bosons) may also
conflict with the non-observation of an antiproton excess in the
PAMELA $\bar{p}/p$ data~\cite{Adriani:2008zq}.

Of course the above observations hold only for a light dark matter
particle of order 100 GeV. For the range of masses expected in our
scenario (around 1 TeV), the proton flux is sufficiently suppressed,
but the positron excess below about 80 GeV would not be explained.
However, the signal can exceed the background at high energy which
would be interesting if experiments can attain the required
sensitivity.

\begin{figure}[t]
\centerline{ 
\includegraphics[width=0.47 \textwidth]{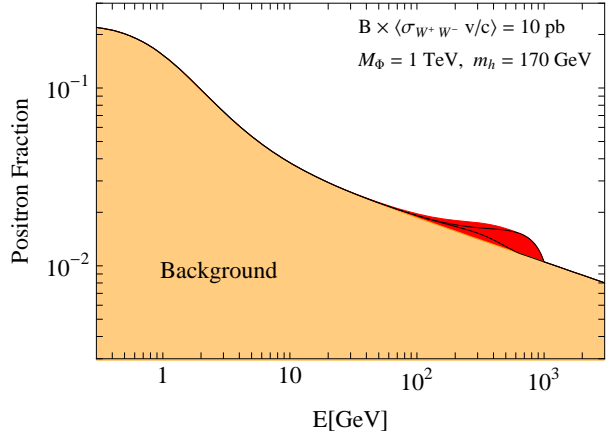}
} 
\caption{Positron fraction when $\Phi\Phi \rightarrow H^{\dagger}H$
(including longitudinal gauge bosons) dominates the DM annihilation
cross section.  We include the secondary positrons/electrons from the
annihilation $\Phi\Phi \rightarrow W^+_{L}W^{-}_{L}$ and the tertiary
positrons/electrons from the annihilation $\Phi\Phi \rightarrow hh
\rightarrow 4W$, assuming $m_{h} = 170~{\rm GeV}$.  Only two-body $W$
decays into positrons/electrons are included.  We take $M_{\Phi} =
1~{\rm TeV}$ and $B \times \langle \sigma_{W^{+}W^{-}} v/c \rangle =
10~{\rm pb}$, with $\rho_{0} = 0.3~{\rm GeV/cm^{3}}$.  The solid lines
represent the individual contributions from secondary and tertiary
production, with the latter having a softer spectrum.}
\label{fig:branepositronexcess}
\end{figure}
We close this subsection by coming back to the case of a scalar DM
candidate that decays dominantly into Higgses, as discussed in
Subsection~\ref{sec:BraneFields}, and comment on the associated
positron signal.  The annihilation cross section
Eq.~(\ref{braneAnnXS}) corresponds to the processes $\Phi\Phi
\rightarrow W_{L}W_{L}$, $\Phi\Phi \rightarrow Z_{L}Z_{L}$ and
$\Phi\Phi \rightarrow hh$ in the large $M_{\Phi}$ limit.  In this
limit we have $\sigma(\Phi\Phi \rightarrow W_{L}W_{L}) \approx
2\sigma(\Phi\Phi \rightarrow Z_{L}Z_{L}) \approx 2\sigma(\Phi\Phi
\rightarrow hh)$.  Further decays of the $W$s, $Z$s and Higgses can
result in energetic positrons.  We consider here the case $m_{H} =
170~{\rm GeV}$ with a SM branching fraction ${\rm BR}(H \rightarrow
W^{+}W^{-}) \approx 1$, and compute the spectrum of \textit{secondary}
positrons from $\Phi\Phi \rightarrow 2W/Z \rightarrow e^{+}X$ and of
\textit{tertiary} positrons from $\Phi\Phi \rightarrow hh \rightarrow
4W \rightarrow e^{+}X$.  We do not include positrons from processes
further down the decay chain.  

The cross sections times branching fractions (including the positron
multiplicities) to be used in Eq.~(\ref{PositronFlux}) are $\langle
\sigma_{W^{+}W^{-}} v \rangle \times {\rm BR}(W\rightarrow e\nu)$, $2
\times \langle \sigma_{ZZ} v \rangle \times {\rm BR}(Z\rightarrow
e^{+}e^{-}) \approx \langle \sigma_{W^{+}W^{-}} v \rangle \times {\rm
BR}(Z\rightarrow e^{+}e^{-})$, and $2 \times \langle \sigma_{HH} v
\rangle \times {\rm BR}(H\rightarrow W^{+}W^{-}) \times {\rm
BR}(W\rightarrow e \nu) \approx \langle \sigma_{W^{+}W^{-}} v \rangle
\times {\rm BR}(W\rightarrow e \nu)$, respectively, where ${\rm
BR}(W\rightarrow e \nu) \approx 0.11$ and ${\rm BR}(Z\rightarrow
e^{+}e^{-}) \approx 0.036$.  In Fig.~\ref{fig:branepositronexcess}, we
show the positron fraction for $M_{\Phi} = 1~{\rm TeV}$ and $B \times
\langle \sigma_{W^{+}W^{-}} v/c \rangle = 10~{\rm pb}$, which
corresponds to a boost factor $B \approx 20$.  We conclude that such a
signal would be visible above background only for rather large boost
factors.  In particular, such a scenario can also not explain the
observed positron excess reported at lower energies.

\section{Conclusions}
\label{sec:conclusions}  

We considered a simple scenario for scalar DM of mass around $1~{\rm
TeV}$ in the context of non-renormalizable theories with a cutoff near
the TeV scale.  Such a possibility arises naturally in
extra-dimensional models that address the hierarchy problem, such as
the Randall-Sundrum scenario but could also arise from other strongly
interacting TeV-scale theories.  The thermal relic density can be
determined either by renormalizable or non-renormalizable
interactions.  Such dark matter particles are clearly more challenging
to detect but can conceivably yield observable gamma ray signals at
current detectors and might ultimately yield observable positrons.

A monochromatic gamma ray line signal arises from the direct
annihilation via nonrenormalizable operators of the DM particle $\Phi$
into photons.  For a cutoff scale of up to about $10~{\rm TeV}$, such
a signal can be larger than the signal from a typical one-loop induced
direct coupling to photons.  We also point out that the monochromatic
signal associated with non-renormalizable operators is likely
observable in currently operating ground-based experiments, and could
be used to probe the cutoff scale up to several TeVs.

We also consider secondary photons from Higgs decay in the
annihilation $\Phi\Phi \rightarrow HH$, and find that this continuous
signal can be observable if the DM halo is relatively peaked at the
galactic center.  Secondary or tertiary positrons can also be produced
in the decays of Higgses or longitudinally polarized gauge bosons, but
the positron flux is likely too small to be observable above
background.

It is also possible to have more exotic scenarios where the
annihilation cross section is dominated by the non-renormalizable
interactions, as opposed to the dimension-4 coupling of $\Phi$ to the
SM Higgs field.  In the extra-dimensional context one could have
annihilations into a SM fermion and the associated KK fermion
dominating the total annihilation cross section.  If the leptonic
channels are dominant, it is possible to have an observable positron
signal in the 100 GeV to 1 TeV range with a boost factor of order one.
Such a signal would typically present two peaks, due to the heavy
lepton involved.  However, the expected flux is too small to account
for the positron excess reported by HESS/AMS-01 and the PAMELA
satellite experiment.  Nevertheless, we find it promising that
indirect searches in the sub-TeV range can be sensitive to cutoff
scale physics.

\bigskip


\subsection*{Acknowledgements}
We would like to thank Douglas Finkbeiner, Fiona Harrison, Sterl
Phinney and Neil Weiner for useful conversations.  E.P. is supported
by DOE under contract DE-FG02-92ER-40699.  L.R. is supported by NSF
grant PHY-0556111.

\appendix

\section{Bulk Scalars in an RS Background}
\label{sec:scalars}  

We consider a bulk real scalar, $\Phi$, propagating in the background
\cite{Randall:1999ee}
\beq
ds^{2} = e^{-2ky} \eta_{\mu\nu} dx^{\mu} dx^{\nu} - dy^{2}~,
\label{lineelement}
\eeq
where $x^{\mu}$ ($\mu = 0,1,2,3$) are the 4D coordinates, and $0\leq y
\leq L$ parametrizes the fifth dimension.  We assume that the scalar
obeys $(-,+)$ boundary conditions, and consider the action
\beqa
S = \int d^{4}x \int^{L}_{0}  \! dy \sqrt{|g|} \, \frac{1}{2} 
\left\{ \partial_{M}\Phi \partial^{M}\Phi - M^{2} \Phi^{2} - 
\delta(y-L) m \Phi^{2} \right\}~,
\eeqa
where $M$ and $m$ are bulk and IR localized mass parameters,
respectively.  We do not write a localized mass on the UV brane, since
the scalar is assumed to vanish at $y=0$.  We will parametrize these
mass parameters in units of the curvature scale as
\beqa
M^{2} &=& \left[c^{2}_{s} + c_{s} - \frac{15}{4} \right] k^{2}~,
\label{Mdef} \\
m &=& \left[c_{s} - \frac{3}{2} + \delta \right] k ~.
\label{mdef}
\eeqa
For $\delta = 0$, the mass spectrum that follows coincides precisely
with that of a fermion obeying $(-,+)$ b.c., where $c_{f} = c_{s}$
parametrizes the fermion bulk mass~\cite{Gherghetta:2000qt}.  Our sign
conventions are such that for $c_{s} < 1/2$ the lightest KK mode is
exponentially localized near the IR brane, and when $c_{s} < -1/2$ its
mass is exponentially smaller than the warped down curvature scale
$\tilde{k} = k \, e^{-kL}$.  We will see that the lightest eigenvalue
can remain small for a wide range of values of the parameter $\delta$
defined in Eq.~(\ref{mdef}), and therefore the lightest scalar KK mode
can be easily lighter than the SM gauge and fermion KK resonances.

The KK decomposition for $\Phi$ reads
\beqa
\Phi(x^{\mu},y) = \frac{e^{ky}}{\sqrt{L}} \sum^{\infty}_{n=1} \phi^{n}(x^{\mu}) f_{n}(y)~,
\label{KKPhi}
\eeqa
where we pulled out an explicit factor $e^{ky}$ for convenience, and
the KK wavefunctions obey
\beqa
\partial^{2}_{y} f_{n} - 2k \partial_{y} f_{n} - (3k^{2} + M^{2}) f_{n} = 
- e^{2ky} m^{2}_{n} f_{n}~,
\eeqa
and satisfy the b.c.:
\beqa
f_{n}(0) |_{y = 0} = 0~,
\hspace{1cm}
\partial_{y} f_{n} |_{y = L} = - (k + m) f_{n}(L)~.
\eeqa
The solutions can be written in terms of Bessel functions as
\beqa
f_{n}(y) = A_{n} e^{ky} \left[ J_{|c_{s} + \frac{1}{2}|} \! 
\left(\frac{m_{n}}{k} e^{ky} \right) 
+ b \, Y_{|c_{s} + \frac{1}{2}|} \! \left(\frac{m_{n}}{k} e^{ky} \right)  
\right]~,
\eeqa
where $A_{n}$ is a normalization constant, determined from
\beqa
\frac{1}{L} \int^{L}_{0} \! dy f_{n}(y) f_{m}(y) = \delta_{nm}~,
\label{KKnormalization}
\eeqa
and
\beqa
b = - \frac{J_{|c_{s} + \frac{1}{2}|} \! \left(\frac{m_{n}}{k} \right)}{Y_{|c_{s} + 
\frac{1}{2}|} \! \left(\frac{m_{n}}{k} \right)}~.
\eeqa
The eigenvalues can be written as $m_{n} = x_{n} k \, e^{-kL}$, where
the $x_{n}$ solve
\beqa
\frac{J_{|c_{s} + \frac{1}{2}|} \! \left(x_{n} e^{-kL} \right)}{Y_{|c_{s} + 
\frac{1}{2}|} \! \left(x_{n} e^{-kL} \right)}=
\frac{x_{n} J_{|c_{s} + \frac{1}{2}| - 1} \! \left(x_{n} \right) + (c_{s} + 
\frac{1}{2} - |c_{s} + \frac{1}{2}| + \delta) J_{|c_{s} + \frac{1}{2}|} \! 
\left(x_{n} \right)}{x_{n} Y_{|c_{s} + \frac{1}{2}| - 1} \! \left(x_{n} \right) + 
(c_{s} + \frac{1}{2} - |c_{s} + \frac{1}{2}| + \delta) Y_{|c_{s} + \frac{1}{2}|} \! 
\left(x_{n} \right)}~.
\label{eigenvalueEqn}
\eeqa
\begin{figure}[t]
\centerline{ 
\includegraphics[width=0.45 \textwidth]{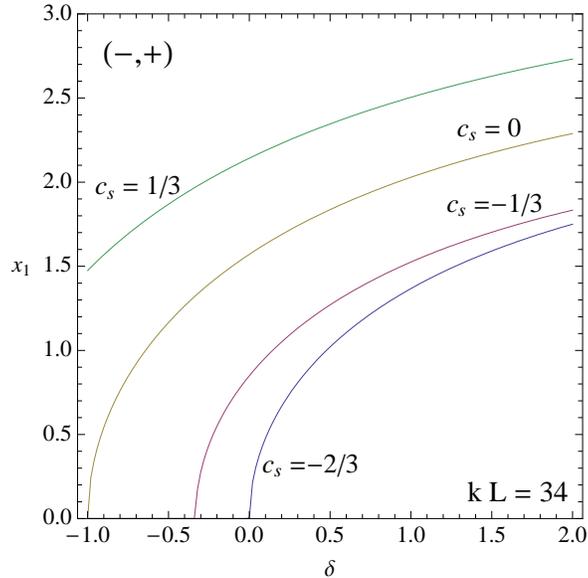}
} \caption{Lightest mass, $m_{1} = x_{1} k \, e^{-kL}$, for a bulk
scalar obeying $(-,+)$ b.c., as a function of bulk and IR localized
masses, parametrized by $c_{s}$ and $\delta$, as in Eqs.~(\ref{Mdef})
and (\ref{mdef}).  Recall that for a KK gauge boson obeying $(+,+)$
b.c., $x_{1} \approx 2.45$.  }
\label{fig:spectrum}
\end{figure}

For $kL \gg 1$, the lowest solutions are approximately given by the
vanishing of the numerator in the r.h.s of Eq.~(\ref{eigenvalueEqn}).
We show the smallest eigenvalue in
Fig.~\ref{fig:spectrum} as a function of $\delta$, defined in
Eq.~(\ref{mdef}), for several values of $c_{s}$ [which parametrizes
the bulk mass $M^{2}$ as in Eq.~(\ref{Mdef})].  This eigenvalue can be
well approximated by
\beqa
x_{1} \approx \left\{ 
\begin{array}{lcl}
2 \sqrt{\frac{1}{2} - c_{s}} \, \sqrt{\frac{\delta}{2+\delta}}  & & c_{s} < -1/2~,   
\\ [0.7em]
2 \sqrt{\frac{3}{2} + c_{s}} \, \sqrt{\frac{1 + 2c_{s} + \delta}{3 + 2 c_{s}+\delta}} 
& & c_{s} > -1/2~.   
\end{array}
 \right.
\label{solx1approx}
\eeqa
We see that for $\delta < \textrm{Min}\{0,-(1+2c_{s})\}$, $x^{2}_{1}$
becomes negative and the corresponding mode is a tachyon.  We will
assume that we are in a region where no such instability arises.  One
should also keep in mind that for $\delta = 0$ and $c_{s} \approx
-1/2$, Eq.~(\ref{solx1approx}) receives additional corrections not
shown there.  In this case, the smallest eigenvalue remains non-zero,
becoming exponentially small for $c_{s} < -1/2$.

For Fig.~\ref{fig:bulkscalar} in the main text, we chose $c_{s} =
-0.2$ and adjusted $\delta$ so as to reproduce the desired mass
$M_{\Phi}$.  This determines the corresponding wavefunction and allows
the computation of the relevant overlap integrals that determine the
$\Phi$ couplings.  Note, however, that the dependence on the choice
$c_{s} = -0.2$ is very mild.


\end{document}